\documentclass[reqno,11pt]{amsart}

\IfFileExists{mymtpro2.sty}{%
  \usepackage[subscriptcorrection]{mymtpro2}
}{}

\usepackage{a4,amssymb}

\usepackage{appendix}

\newtheorem{theorem}{Theorem}[section]
\newtheorem{lemma}{Lemma}[section]

\newtheorem{proposition}{Proposition}[section]
\newtheorem{remark}[theorem]{Remark}
\newtheorem{remarks}[theorem]{Remarks}

\newtheorem{definition}[theorem]{Definition}
\newcommand{\labelnummer}{\mbox{\normalfont (\roman{numcount})}}%

\makeatletter

  {\let\curlabelspeicher\@currentlabel%
    \begin{list}{\labelnummer}%
      {\usecounter{numcount}\leftmargin0pt%
        \topsep0.5ex\partopsep2ex\parsep0pt\itemsep0ex\@plus1\p@%
        \labelwidth2.5em\itemindent3.5em\labelsep1em%
      }%
    \let\saveitem\item%
    \def\item{\saveitem%
      \def\@currentlabel{{\upshape\curlabelspeicher}$\,$\labelnummer}}%
    \let\savelabel\label%
    \def\label##1{\savelabel{##1}%
      \@bsphack%
        \ifmmode\else%
          \protected@write\@auxout{}%
          {\string\newlabel{##1item}{{\labelnummer}{\thepage}}}%
        \fi%
      \@esphack%
    }%
  }{\end{list}}%

\renewcommand{\appendix}{\def\thesection{\textsc{Appendix}}}

 \let\leq\le
 \let\geq\ge

 \let\Im\undefined

\DeclareMathOperator{\Im}{Im}

\DeclareMathOperator{\tr}{tr\kern1pt}

\makeatletter

\newif\ifper\pertrue
\def\per{.}

\def\bti{\@ifnextchar[\bbti\bbbti}
\def\bbti[#1]#2{#2, #1.}
\def\bbbti#1{#1.}

\def\z{\@ifnextchar[\zz\zzz}
\def\zz[#1]#2#3#4#5{\perfalse\emph{#2} \textbf{#3}, #4 (#5) [#1]}
\def\zzz#1#2#3#4{\emph{#1} \textbf{#2}, #3 (#4)\ifper\per\fi\pertrue}

\def\pub{\@ifstar\pubstar\pubnostar}
\def\pubnostar{\@ifnextchar[\@@pubnostar\@pubnostar}
\def\@@pubnostar[#1]#2#3#4{#2, #3, #4, #1\ifper\per\fi\pertrue}
\def\@pubnostar#1#2#3{#1, #2, #3\ifper\per\fi\pertrue}
\def\pubstar[#1]#2#3#4{\perfalse #2, #3, #4 [#1]\pertrue}

\makeatother

\newcommand{\beq}{\begin{equation}}
\newcommand{\eeq}{\end{equation}}
\newcommand{\ba}{\begin{array}}
\newcommand{\ea}{\end{array}}
\newcommand{\bea}{\begin{eqnarray}}
\newcommand{\eea}{\end{eqnarray}}

\newcommand{\R}{\mathbb{R}}

\newcommand{\Z}{\mathbb{Z}}

\newcommand{\N}{\mathbb{N}}

\newcommand{\C}{\mathbb{C}}

\DeclareMathOperator{\supp}{\mathrm{supp}}

{

\newcommand{\Schr}{Schr\"odinger }

\def\P{I\kern-.30em{P}}
\def\E{I\kern-.30em{E}}
\renewcommand{\E}{\mathbb{E}\mkern2mu}
\renewcommand{\P}{\mathbb{P}}

\begin{document}

\title[Local eigenvalue statistics]{Local eigenvalue statistics for higher-rank Anderson models after Dietlein-Elgart}

\author[S.\ Herschenfeld]{Samuel Herschenfeld}
\address{Department of Mathematics,
    University of Kentucky,
    Lexington, Kentucky  40506-0027, USA}
\email{samuel.herschenfeld@uky.edu}

\author[P.\ D.\ Hislop]{Peter D.\ Hislop}
\address{Department of Mathematics,
    University of Kentucky,
    Lexington, Kentucky  40506-0027, USA}
\email{peter.hislop@uky.edu}

\thanks{PDH is partially supported by Simons Foundation Collaboration Grant for Mathematicians No.\ 843327.  This article is partially based on the doctoral dissertation of SH submitted in partial fulfillment of the PhD degree in mathematics at the University of Kentucky.}


\begin{abstract}
We use the method of eigenvalue level spacing developed by Dietlein and Elgart \cite{de} to prove that the local eigenvalue statistics (LES) for the Anderson model on $\Z^d$, with uniform higher-rank 
$m \geq 2$, single-site perturbations, is given by a Poisson point process with intensity measure $n(E_0)~ds$, where $n(E_0)$ is the density of states at energy $E_0$ in the region of localization near the spectral band edges. This improves the result of Hislop and Krishna \cite{hk2015}, who proved that the LES is a compound Poisson process with L\'evy measure supported on the set $\{1, 2, \ldots, m \}$.
Our proofs are an application of the ideas of Dieltein and Elgart to these higher-rank lattice models with two spectral band edges, and illustrate, in a simpler setting, the key steps of the proof of Dieltein and Elgart.
\end{abstract}

\maketitle \thispagestyle{empty}

\tableofcontents



\section{Statement of the problem}\label{sec:introduction}
\setcounter{equation}{0}

A long outstanding problem in the theory of random \Schr operators (RSO) was the extension of the Minami estimate from RSO on the lattice $\Z^d$ to those on the continuum $\R^d$.  The Minami estimate was introduced for lattice Anderson model (projections with rank $m = 1$) by Minami \cite{minami1} (see also Bellissard, Hislop, and Stolz \cite{bhs} and Graf and Vaghi \cite{gv}) in order to prove that the local  eigenvalue statistics (LES), for energies in the  localization regime, is given by a Poisson point process. The Minami estimate also plays a key role in the proof of the simplicity of eigenvalues in the localization regime and the asymptotic independence of local point processes associated with disjoint energy intervals (see \cite{hks} and \cite{klopp2}). For RSO on $L^2 (\R)$, a Minami estimate was proven by Klopp \cite{klopp1} using localization and a Wegner estimate, see also Herschenfeld \cite[Chapter 3]{herschenfeld_thesis}. Analogous results for RSO on $\ell^2(\Z)$, obtained using techniques similar to \cite{klopp1}, were proven by Shirley \cite{shirley1,shirley2}. 

In a recent paper \cite{de}, Dietlein and Elgart proved a weak Minami estimate for RSO on $L^2(\R^d)$ for energies in a small interval near the bottom of the deterministic spectrum. The simplicity of eigenvalues and the Poisson nature of the local point processes then follow from the weak Minami estimate and the ideas in \cite{cgk2,kleinMolchanov}. The basic model treated in \cite{de} is the continuum Anderson model for which $H_\omega := - \Delta + V_\omega$, acting on $L^2 (\R^d)$. The random potential has the form
\beq\label{eq:pot1}
V_\omega(x) := \sum_{k \in \Z^d} \omega_k u(x-k) ,
\eeq
where the single-site potential satisfies $u \geq 0$, ${\rm supp} ~u$ is compact and contains neighborhood of the unit cube, and for some $\delta >0$, the covering condition $0 < \delta \leq\sum_k u(x-k) < \infty$ is satisfied.
The random variables $\{ \omega_j \}$ are independent and identically distributed with values in $[0,1]$. With these assumptions, the deterministic spectrum $\Sigma = [ 0 , \infty)$. (This normalization is made here for convenience.)

One of the novel ideas in \cite{de} is to consider the eigenvalue level spacing function (EVLS) defined as follows. For a self-adjoint Hamiltonian $H$ with discrete spectrum $\{ E_j \}$, listed with multiplicity, the EVLS  of $H$ in interval $I \subset \R$ is defined by
\beq\label{eq:evls_defn1}
{\rm spac}_I (H) := \min_{ j \neq k } \{ | E_j - E_k | ~|~  E_j , E_k \in \sigma(H) \cap  I \}.
\eeq
This notion is convenient since if ${\rm spac}_I (H)  > \delta$, then for any $E+(-\frac{\delta}{2},\frac{\delta}{2})=:I_{\delta,E}\subset I$, $H$ has at most one eigenvalue in $I_{\delta,E}$.  In particular, if ${\rm spac}_I (H)  > 0$, then all the eigenvalues of $H$ in $I$ are nondegenerate.  

One can outline the Dietlein-Elgart approach to a weak Minami estimate by way of the EVLS for a \Schr operator $H_\omega^L := H_\omega | \Lambda_L$, localized to a cube $\Lambda_L \subset \R^d$ with Dirichlet boundary conditions, as follows:

 \begin{enumerate}
 
 \item \emph{Perturbation theory and good configurations}: Suppose for some fixed configuration $\omega_0 : \{ (\omega_0)_j ~|~ j \in \Z^d \}$, the local Hamiltonian $H_{\omega_0}^L$ has $n$ eigenvalues in an interval $I \subset [0,E_{\text{sp}}]$, with $E_{\text{sp}}\approx \frac{\pi^2}{2}$, the first nonzero Neumann eigenvalue of the Laplacian on a unit cube, contained in the support of $u$. We also suppose that $I$
is isolated from the rest of the spectrum of $H_{\omega_0}^L$. 
Then, there is a good configuration, $\widetilde{\omega}$, near $\omega_0$, for which there is a positive lower bound on the EVLS of $H_\omega^L$. That is, in an $\epsilon$-neighborhood of any such initial configuration $\omega_0$, there is a good configuration $\widetilde{\omega}$, so that $H_{\widetilde{\omega}}^L$ has $n$-eigenvalues in $I_{{\epsilon}}:=I+[-\epsilon,\epsilon]$, and its EVLS satisfies
 $$
 {\rm spac}_{I_\epsilon} (H_{\widetilde{\omega}}^L)  > \delta(\epsilon) ,
 $$
where $\delta(\epsilon)\sim \epsilon L^{-\alpha}$, $\alpha>0$.

%

%
 
  \item \emph{Eigenvalue level spacing estimate (EVLS) and bad configurations}: The perturbation result of step 1 is useful only if there are sufficiently many good configurations. A key result of Dietlein and Elgart is an estimate on the size of the bad configurations. This is the probabilistic bound that 
 $$
 \P \{ {\rm spac}_{I_E} (H_\omega^L) < \delta \} \leq C_1 L^{2d} | \log \delta|^{-K} ,
 $$ 
 where $I_E := [0, E]$, for any $E \in (0, E_\text{sp})$.  
The proof is based on the relationship between the EVLS function $F(\omega) :=  {\rm spac}_{I_E} (H_\omega^L)$, and the discriminant $G(\omega)$, defined by
$$
G(\omega) := {\rm disc}_{I_E} (H_\omega^L) := \prod_{1 \leq j < \ell \leq n} (E_j^L (\omega) - E_\ell^L (\omega))^2. 
$$
The advantage of $G(\omega)$ is that it is locally analytic in $\omega$. 
The measure of the set of bad configurations for which the EVLS $F(\omega)$ is small
can be estimated using $G(\omega)$ and    
a Cartan-type lemma. This lemma provides an upper bound on the size of the set of nearby bad configurations. 
 
 \medskip
 
 \item \emph{A weak Minami estimate}: The EVLS estimate is a key ingredient in the proof of a version of Minami's original estimate, valid for energies $E$ near the bottom of the deterministic spectrum:
 $$
  \P \{ {\rm Tr}_{E+[-\frac{\delta}{2},\frac{\delta}{2}]} (H_\omega^L) \geq 2 \} \leq C_M  L^{4d} \delta | \log \delta|^{-K} ,
 $$ 
 for any $K > 0$ and where $C_M$ depends on $K$. 
 \end{enumerate}
We mention that localization is not used in any of these steps.  The estimate in (3) is a weaker 
probability bound than Minami's original estimate, but is sufficient to prove that the LES  is given by a Poisson point process and the simplicity of the point spectrum. Localization bounds are needed for these proofs, see also remark \ref{rmk:loc1}.

The purpose of this article, part pedagogical and part new, is to show how the methods of Dietlein and Elgart \cite{de} apply to the higher-rank Anderson model on the lattice as described in \cite{hk2015}. The standard proof of the Minami estimate, see, for example \cite{minami1} and \cite{cgk1},  does not apply to this model. An additional complication arises since lattice polymer models
have a deterministic spectrum with upper and lower band edges. We show how the methods of Dietlein-Elgart apply to neighborhoods of both band edges. Additionally, the higher-rank Anderson model allows for some simplifications that, perhaps, more clearly illustrate the fundamental ideas of Dietlein and Elgart. In section \ref{subsec:simplifications1}, we indicate certain simplifications that occur for the higher-rank Anderson models, and how the arguments have to be modified for \Schr operators on $L^2 (\R^d)$ as in \cite{de}.   


The second author and M.\ Krishna \cite{hk2015} proved that the LES for the higher-rank Anderson model is given by a compound Poisson point process with finitely supported Levy measure.  An extension of the method of Dietlein-Elgart, presented in this paper, allows us to prove that, in fact, the process is a Poisson point process (in a smaller interval than in \cite{hk2015}) The extension consists of treating both of the band edges of the almost sure spectrum in a uniform manner. The proof of the EVLS and the Minami estimate for this model is more complicated than the rank-one Anderson model \cite{minami1}, since the standard proofs do not work, but is less complex than for the continuum model since it allows for certain simplifications. We believe that the higher-rank Anderson model presents a nice, nontrivial illustration of the method of \cite{de}.

 
%

 
 \subsection{The higher-rank Anderson model}\label{subsec:model1}
 
 We consider a discrete random Schr\"odinger operator $H_\omega := H_0 + V_\omega$ of alloy-type  on the Hilbert space $\ell^2(\Z^d)$.  The fixed operator $H_0$ is the positive, discrete Laplacian,
\begin{equation}\label{eq:disc_laplace1}
(H_0 f)(n) = 2df(n) - \sum_{k:|k-n|=1}f(k).
\end{equation}
By taking the Fourier transform of $H_0$ in \eqref{eq:disc_laplace1}, it is easily seen that the spectrum, denoted by  $\sigma(H_0)$, is $\sigma(H_0)=[0,4d]$.

The rank-one Anderson-type potential $V_\omega$ is a random potential 
\beq\label{eq:potRank1}
V_\omega  := \sum_{k \in \Z^d} \omega_k P_k, 
\eeq
where $P_k = | k \rangle \langle k|$, for $k \in \Z^d$, is the rank-one projection on site $k \in \Z^d$.
The family of coefficients $\{\omega_k\}_{k\in \Z^d}$ is a collection of independent, identically distributed (iid), bounded, random variables with continuous density $\rho$ and $\supp \rho =[0,1]$.
%

In this article, we are interested in a higher-rank version of this model defined as follows. 
Let $\Lambda_r(n)=\prod_{j=1}^d \{n_j,\cdots,n_j+r-1\}$ be a cube with side length $r\in \N$ and vertex at $n\in \Z^d$.  Let $\chi_\Lambda$ be the characteristic function on $\Lambda$, or equivalently in this model, $\chi_\Lambda=P_\Lambda$, the projection onto the sites in $\Lambda$. The projection $P_{\Lambda_r (n)}$ has rank $|\Lambda_r(n)|=r^d \geq 1$. The analog of \eqref{eq:potRank1} for a rank-$r^d$ Anderson-type random potential is, 
\beq\label{eq:potRankM}
V_\omega  := \sum_{k \in r \Z^d} \omega_k P_k, 
\eeq
for which $P_k$ is a rank-$r^d$ projection. For notational simplicity, we set $m :=r^d$, the uniform rank of the projections $P_k, k \in r \Z^d$. 

We assume that the common density $\rho$ of the random variables $\omega_j$ 
is Lipschitz continuous with $\supp \rho =[0,1]$:  
\begin{equation}
|\rho(x)-\rho(y)|<\mathcal{K}|x-y| \text{ and } 0 < \rho_-<\rho(x)<\rho_+
\end{equation}
for some $\mathcal{K},\rho_-,\rho_+\in (0,\infty)$ and all $x,y\in[0,1]$.

It is well-know that for a family of ergodic, random \Schr operators, there exists a closed subset $\Sigma \subset \R$ such that  $\sigma(H_\omega)=\Sigma$, for almost every configuration of random variables $\omega := \{ \omega_j \}_{j \in r \Z^d }$. This set $\Sigma$ is called the deterministic spectrum of the family. 
The assumptions on our model imply that the deterministic spectrum of the family $\{ H_\omega ~|~ \omega \in \Omega \}$ is,
\begin{equation}
\Sigma=\sigma(H_0)+\mathrm{supp}(\rho) = [0,4d+1].
\end{equation}

We now introduce the finite-volume \Schr operators $H_0^L$ and  $H_\omega^L$. These operators are the restrictions of $H_0$ and $H_\omega$, respectively, to bounded cubes $\Lambda_L$ for the specific choice of $L \in r\Z$. 
For $L\in r\Z$, we choose $\Lambda_L$ to be a cube of side length $L$ such that for an index set $\Lambda_{r,L}^*\subset r \Z^d$ with $|\Lambda_{r,L}^*| = ({\frac{L}{r}})^d$, we have (up to sets of measure zero coming from the boundaries)
\beq\label{eq:decomp1}
\Lambda_L =\bigcup_{k\in \Lambda_{r,L}^*} \Lambda_r(k) .
\eeq
We define the local free Hamiltonian $H_0^L$ to be the restriction of $H_0$ to $\Lambda_L$ so that $H_0^L = \chi_{\Lambda_L} H_0 \chi_{\Lambda_L}$. This corresponds to \emph{simple boundary conditions}. We will also need restrictions of $H_0$ to $\Lambda_L$ with Dirichlet and Neumann boundary conditions, see Appendix \ref{app3:bc_free1} for a review of these operators. 
The local potential is defined by 
$$
V_\omega^L := \sum_{k\in \Lambda_{r,L}^*} \omega_k \chi_{\Lambda_r(k)} = \chi_{\Lambda_L}V_\omega  .
$$
Consequently, the local \Schr operator $H_\omega^L$, with simple boundary conditions, has the form
\beq\label{eq:localSchr1}
H_\omega^L := H_0^L + V_\omega^L ,
\eeq
acting on $\ell^2 (\Lambda_L)$. 


\subsection{Results for higher-rank Anderson models}\label{subsec:main_thms1}

The higher-rank Anderson model allows us to illustrate the methods of Dietlein-Elgart in a simpler setting than the continuum model and also to extend the results to this model with lower and upper band edges. 
Based on the methods of Dietlein and Elgart \cite{de}, we prove the following results, for energy intervals near each band edge, for the higher-rank Anderson model:

\begin{itemize}
\item Eigenvalue level spacing theorem, Theorem \ref{thm:spacing}; 

\medskip
\item weak Minami estimate, Theorem \ref{thm:minami1};

\medskip
\item Almost sure simplicity of the eigenvalues, Theorem \ref{thm:simplicity1};

\medskip
\item Poisson nature of the local eigenvalue statistics, Theorem \ref{thm:poisson1}.

\end{itemize}


\noindent
In more detail, this model allows us to  
\begin{enumerate}
\item Illustrate the method of Dietlein-Egart by applying it to an interesting lattice model that cannot be treated by other methods. The model allows for simplifications of some of the technicalities necessary for the continuum model so that the key ideas are more transparent;

\medskip

\item Apply the method of Dietlein-Elgart to discrete \Schr operators at both the lower and upper band edges.
\end{enumerate}


%
%
%
%
%
%
%
%
%
%
%



\subsection{Simplifications for the higher-rank Anderson model}\label{subsec:simplifications1}

The higher-rank Anderson model on $\ell^2 (\Z^d)$ brings several simplifications to the Dietlein-Elgart argument compared to the treatment of \Schr operators on $L^2 (\R^d)$. In particular, we note the following:

\begin{enumerate}
\item The covering condition. This is the requirement that the translations of a single-site potential $u \geq 0$ satisfy $\sum_{k \in \Z^d} u(x-k) \geq \delta > 0$, for all $x \in \R^d$. This is satisfied for the higher-rank Anderson models since \eqref{eq:potRankM} implies that $\sum_{k \in rZ^d} P_k = I$, the identity on $\Z^d$. 

\medskip

\item The need for the modified operators $-G \Delta G$ and $V^{-\frac{1}{2}}(H_\omega - E)V^{- \frac{1}{2}}$ is related to the covering condition. In the proof of the Minami-type estimate in \cite[section 5]{de}, Dietlein and Elgart use a crucial identity: A uniform shift in the random variables $\omega_k + \tau$, for all $k \in \Lambda_L \cap \Z^d$, results in a shift in the energy: $H_{\omega + \tau}^L = H_\omega^L + \tau \chi_{\Lambda_L}$. Since $H_{\omega + \tau}^L = H_0^L + V_{\omega + \tau}^L$, and 
$$
V_{\omega + \tau}^L (x) = \sum_{k \in \Lambda_L}( \omega_k + \tau) u(x-k) =
 V_\omega^L(x) + \left(\sum_{k \in \Lambda_L} u(x-k) \right) \tau , 
 $$ 
the condition $H_{\omega + \tau}^L = H_\omega^L + \tau \chi_{\Lambda_L}$ requires that 
\beq\label{eq:cover1}
V_L(x) := \left(\sum_{k \in \Lambda_L} u(x-k) \right) = \chi_{\Lambda_L} .
\eeq
Even though the covering condition guarantees that  
$V_L(x) \geq \delta \chi_{\Lambda_L}$, this lower bound is not sufficient. By replacing $H_\omega^L$ with $V_L^{-1/2} H_\omega^L V_L^{-1/2}$, the new potential will satisfy condition \eqref{eq:cover1}. For this reason, in sections $1$ through $4$ on the EVLS theorem, Dietlein and Elgart treat the more general operators $- G \Delta G$, for a smooth function $G$ bounded from below $G \geq G_- > 0$. This condition is automatically satisfied for the higher-rank Anderson models and the calculations in section \ref{sec:minami1} are simplified.

\medskip

\item Special Wegner estimate, \cite[Lemma 4.4]{de}. Another advantage of the higher-rank Anderson model is that there is a generalized Minami estimate, Theorem \ref{thm:gMinami}.  This allows us to avoid the use of the discrete analog of  \cite[Lemma 4.4]{de} which, however, is of interest in its own right.

\medskip

\item The proof of  \cite[Lemma 4.4]{de} uses estimates on the spectral shift function (SSF) generalizing the rank-$m$ perturbation bound for lattice models: 
$$
|\tr \chi_I(H_\omega^L) - \tr \chi_I(H_{\omega_n^\perp,\widetilde{\omega}_n}^L) | \leq m .
$$   
This bounds the number of eigenvalues potentially perturbed by a perturbation $\omega_n \rightarrow \widetilde{\omega}_n$ of rank $m$. 

\end{enumerate}

\subsection{Some open problems}

The work of Dietlein and Elgart has opened many interesting paths of research. Some of the open problems of interest include:

\begin{enumerate}
\item Extensions of the method to \Schr operators with multiple band edges for both lattice and continuum models;

\medskip

\item Extensions of the method to continuum random \Schr operators with a constant magnetic field in two-dimensions;

\medskip

\item Extending the technique to larger energy intervals, for example, to the entire localization regime. It should be noted that the Minami estimate for the rank-one Anderson model is valid throughout the entire spectrum. One of the current limitations of the method comes primarily from the lower bound estimate based on the first nonzero Neumann eigenvalue. 

\medskip

\item Elimination of the covering condition for the potential for continuum models. This has been done for the Wegner estimate and perhaps similar techniques will apply to the Dietlein-Elgart method of proof of the Minami estimate (see Remark \ref{rmk:covering2}). 

\end{enumerate}

%

\subsection{Contents of the paper}\label{subsec:contents}

 The rest of this paper is organized as follows.  
 In section \ref{sec:background1}, we give a brief review of some concepts for the rank-one Anderson model and indicate how the classic proof of the Minami estimate for the rank-one Anderson model does not extend to higher-rank models. The generalized Minami estimate for higher-rank models is also proved. Section \ref{sec:evDeg1} is devoted to a study of the eigenvalue level spacing. Perturbation theory is used to establish the existence of good configurations for which the degeneracies of all eigenvalues in an isolated cluster of eigenvalues are removed. The main EVLS theorem is proven in section \ref{sec:evls1}, including a the key probabilistic estimate, based on the Cartan-like lemma, on the size of the set of bad configurations. The first main result, a weak Minami estimate, is proven in section \ref{sec:minami1}. The two  main applications are described and proved in section \ref{sec:simplicity1}, the simplicity of the eigenvalues, and in section \ref{sec:PPP1}, the Poisson nature of the LES.  Model-independent functional analytic tools necessary for these proofs are described in appendices \ref{app1:FunctAnaly1} and \ref{app2:cartan}. Appendix \ref{app3:bc_free1} contains a description of some of the self-adjoint boundary conditions for the Laplacian restricted to regions in $\Z^d$, and some spectral properties of lattice operators, used in section \ref{sec:evDeg1}.





\section{Background: Wegner and generalized Minami estimates for higher-rank Anderson models}\label{sec:background1}
\setcounter{equation}{0}

In this section, we discuss the two main estimates for lattice Anderson-type Hamiltonians restricted to finite cubes in $\Z^d$. The Wegner estimate is an upper bound on the probability that the local Hamiltonian $H^L_\omega$ has at least one eigenvalue in an interval $I \subset \R$. It provides an upper bound on the density of states measure of the interval $I$. As we will show, the estimate has the same form for rank-one and higher-rank Anderson models. 

In general, a Minami estimate  is
 an upper bound on the probability that several eigenvalues of $H_\omega^L$ lie in an interval $I \subset \R$, and hence involves correlations between eigenvalues.  For the rank-one case, the classic Minami estimate gives an upper bound on the probability that there are two or more eigenvalues in an interval $I$. However, for the uniform rank-$m>1$ case, a only generalized Minami estimate holds for the probability that there are $m$ or more eigenvalues in an interval $I$. 
The proof of this generalized Minami estimate is based on spectral averaging and presented in Theorem \ref{thm:gMinami}. 

\subsection{Spectral averaging and the Wegner estimate}\label{subsec:wegner1}

 The Wegner estimate is a first result on the distribution of the eigenvalues of the the local \Schr operator $H_\omega^L$. 
It is an essential ingredient in the proofs of localization.  It also provides context for understanding the Minami estimate and is used in the proof of the Minami estimate.  We include here a proof of the Wegner estimate for the higher-rank Anderson model  based on the proof for the continuum models in \cite[Lemma 4.1, Corollary 4.2]{ch94}. The standard proof for the rank-one Anderson model 
\cite{cgk1} seems not to be applicable to the rank-$m$ models considered here. 

 Spectral averaging is an important ingredient in the proof of the Wegner estimate presented here.
 The abstract version of spectral averaging states that for certain measures, depending on a real parameter, the measure obtained by averaging these measures with respect to this parameter yields a measure absolutely continuous with respect to Lebesgue measure. For our application, the parameter is a single random variable, say $\omega_k$, and we will average diagonal matrix elements of the resolvent with respect to $\omega_k$.  We denote this averaging by $\E_{\omega_k}$. 



We begin with a review of spectral averaging in the  rank-one case. 
The perturbation associated to a single random variable $\omega_k$, for $k \in \Lambda_L$, is linear in 
$\omega_k$, and we may write $H_\omega^L = H_{\omega_k^\perp}^L + \omega_k P_k$, where
the configuration $\omega = ( \omega_k, \omega_k^\perp )$, and $P_k$ is a rank-one projection onto site $k \in \Lambda_L$.

\begin{proposition}[Spectral averaging-rank-$1$]\label{prop:specAve_r1}
For any $z \in \C^+$, we have  
\beq\label{eq:specAve01}
\int \rho(\omega_k) \Im \langle \delta_{k}, R_\omega^L (z) \delta_{k} \rangle ~d \omega_k \leq 
\pi \| \rho \|_\infty .
\eeq
Consequently, for any interval $I \subset \R$, we have
\beq\label{eq:specAve02}
\E_{\omega_k} \{ \langle \delta_k, \chi_I(H_\omega^L) \delta_k \rangle \} \leq  \| \rho \|_\infty |I|. 
\eeq
\end{proposition}

\begin{proof}
With respect to the perturbation $\omega_k P_k$, the second resolvent formula gives 
\beq\label{eq:specAve1}
\langle \delta_{k}, R_{\omega_k}^L (z) \delta_{k} \rangle = \frac{1}{ \omega_{k} + \langle \delta_{k}, R_{H_{\omega_k^\perp}^L}  (z) \delta_{k} \rangle^{-1} },
\eeq
We now compute the imaginary part of the matrix element that is independent of $\omega_k$. We write $\langle \delta_{k}, R_{\omega_k^\perp}^L (z) \delta_{k} \rangle ^{-1} := f_{Re}(z) + i f_{Im}(z)$, for real-valued functions $f_{Re}$ and $f_{Im}$, independent of $\omega_k$, so that 
\beq\label{eq:specAve2}
\Im \langle \delta_{k}, R_{\omega_k}^L (z) \delta_{k} \rangle   = \frac{f_{Im (z)}}{ (\omega_k + f_{Re}(z) )^2 + f_{Im}(z)^2 } .
\eeq
Integrating this expression with respect to $\omega_k$ with measure $\rho(\omega_k) ~d\omega_k$, yields
\beq\label{eq:specAve3}
\int \rho(\omega_k) \Im \langle \delta_{k}, R_\omega^L (z) \delta_{k} \rangle ~d \omega_k \leq 
\pi \| \rho \|_\infty .
\eeq
The second result \eqref{eq:specAve02} follows from an application of Stone's formula (see \eqref{eq:stone1}). 
\end{proof}

Spectral averaging is a key ingredient in the the proof of the Wegner estimate. This is an upper bound on the probability that the local Hamiltonian $H_\omega^L$ has at least one eigenvalue in an interval $I$. 
 
\begin{theorem}[Wegner Estimate-rank $1$]\label{thm:Wegner}
Let $H_\omega^L$ be the local rank-one Anderson model defined in 
\eqref{eq:localSchr1} with random variables satisfying the above assumptions. 
Then, for any interval interval  $I  \subset \R$, we have
\begin{equation}\label{eq:wegner_r1}
\mathbb{P} \{ \chi_I(H_\omega^L) \geq 1 \} \leq \E \{ \chi_I(H_\omega^L) \} \leq  \|\rho\|_\infty |\Lambda_L| |I| .
\end{equation}
\end{theorem}

\begin{proof}
The independence of the random variables $\{ \omega_k \}_{k \in \Lambda_,L}$ 
allows us to write for any random variable $X = X( \omega)$
$$
\mathbb{E} \{ X \} =\mathbb{E}_{\omega_k^\perp}\mathbb{E}_{\omega_k} \{ X \}, 
$$
for any $k \in \Lambda_{L}$.  We write $\mathbb{E}_{\omega_k}$ to denote the expectation with respect to just $\omega_k$, and $\mathbb{E}_{\omega_k^\perp}$ to denote the expectation with respect to all other variables, $\{\omega_m\}_{m\neq k}$, for $m \in \Lambda_{L}$.
The Chebychev inequality, the decomposition of the expectation, and spectral averaging imply
\bea\label{eq:wegner1}
\mathbb{P} \{ \tr \chi_I(H_\omega^L)\geq 1 \} & \leq & \mathbb{E} \{ \tr\chi_I(H_\omega^L) \}  \nonumber \\
 & = & \sum_{k \in \Lambda_{L}} \mathbb{E}_{\omega_{k^\perp}} \mathbb{E}_{\omega_k}  \{ 
 \langle \delta_k, \chi_I(H_\omega^L) \delta_k \rangle  \} \nonumber  \\
  & \leq & \|\rho\|_\infty |\Lambda_L| |I|,
    \eea
  by \eqref{eq:specAve02}.  
\end{proof}

When the projectors $P_k$ are of rank $m > 1$, the second resolvent formula no longer yields the simple expression resulting in \eqref{eq:specAve1} since off-diagonal terms arise from the term $R_{H_\omega^L} (z) P_k R_{H_{\omega_k}^L}(z)$. 
Instead, spectral averaging for the higher-rank Anderson model may be proved using the ideas of \cite[Lemma 4.1, Corollary 4.2]{ch94}. 

\begin{proposition}[Spectral averaging-rank $m$]\label{prop:specAve1}
For any interval  $I \subset \R$, there is a constant $C_W > 0$, depending only on the support of $\rho$,  
so that for any $k \in \Lambda_{L,r}^*$, and for any $\varphi \in \ell^2( \Z^d)$, with $\| \varphi \| = 1$, we have 
\beq\label{eq:spec_ave1}
 \E_{\omega_k} \{ \langle \varphi,  P_k  \chi_I (H_{(\omega_k, \omega_k^\perp)}^L) P_k  \varphi \rangle \}  \leq C_W  
\| \rho \|_\infty  |I| .
\eeq
Consequently,  we have the estimate
\beq\label{eq:spec_ave2}
\E_{\omega_k} \{ {\rm tr} P_k \chi_I(H_{(\omega_k, \omega_k^\perp)}^L ) P_k \} \leq m C_W  
\| \rho \|_\infty  |I| .
\eeq
\end{proposition}

\begin{proof}
1. We write $H_{\omega_k}^L$ for $H_{(\omega_k, \omega_k^\perp)}^L$, since $\omega_k^\perp$ will be held fixed in this calculation. Because $H_{\omega_k}^L$ is linear in $\omega_k$, we can extend $\omega_k$ to complex values. 
We define a function $K(\lambda, z) := P_k (H_{(\lambda, \omega_k^\perp)}^L - z)^{-1} P_k$, for any complex $\lambda$ with $\Im \lambda > 0$ and $\Im z < 0$. 
Using the second resolvent formula, we easily derive a key inequality: 
\beq\label{eq:key_identity1}
- \Im K(\lambda,z) = ( \Im \lambda - \Im z) K^*(\lambda,z) K(\lambda, z) \geq  (\Im \lambda)  K(\lambda,z)^*  K(\lambda,z),
\eeq
that is easily derived using the second resolvent formula. As a result, we have,
\beq\label{eq:key_identity2}
\| K(\lambda, z) \| \leq \min \left( \frac{1}{\Im \lambda} , \frac{1}{| \Im z | } \right) .
\eeq
For $t > 0$, $\epsilon \geq 0$, and $\Im z < 0$, we define a function $F_t(\epsilon, z)$ by
$$
F_t(\epsilon,z) 
:= \int_\R \frac{1}{1 + t \lambda^2}  K(\lambda + i \epsilon,z) ~d \lambda .
$$
Due to the analyticity of $K(\lambda,z)$ with respect to $\lambda$ in the upper-half complex plane, provided $\Im z < 0$, the integral may be evaluated by residues resulting in
\beq\label{eq:key_identity3}
  F_t(\epsilon,z) = \frac{\pi}{t^{1/2}} K(i(t^{- 1/2} + \epsilon), z) ,
  \eeq
 so that by \eqref{eq:key_identity2} it follows that 
 \beq\label{eq:key_identity4}
 \| F_t(\epsilon,z)  \| \leq \frac{\pi}{t^{1/2}} \min \left( \frac{t^{1/2}}{1 + t^{1/2} \epsilon} , \frac{1}{| \Im z | } \right).
 \eeq
For any given $z$, we may choose $\epsilon > 0$ and $t >0$ so that $(|\Im z| - \epsilon) t^{1/2} < 1$. 
It follows from \eqref{eq:key_identity4} that 
\beq\label{eq:key_identity5}
 \| F_t(\epsilon,z)  \| \leq {\pi},
 \eeq
 uniformly with respect to $z \in \C^-$ and $\epsilon \geq 0$. 
 
\noindent
2. By Stone's formula, we have for any $\varphi \in \ell^2 ( \Z^d)$, with $\| \varphi \| = 1$, 
\beq\label{eq:stone1}
\langle \varphi, P_k \chi_I (H_{\omega_k}) P_k \varphi \rangle = -\frac{1}{\pi} \lim_{\delta \rightarrow 0} \int_I  \Im \langle \varphi, P_k (H_{\omega_k} - E +i \delta)^{-1} P_k  \varphi \rangle ~dE .
\eeq
Taking the expectation with respect to $\omega_k$ yields
\bea\label{eq:stone2}
\lefteqn{\E_{\omega_k}  \{ \langle \varphi, P_k \chi_I (H_{\omega_k}) P_k \varphi \rangle \} } \nonumber \\
 & = &
- \frac{1}{\pi} \lim_{\delta \rightarrow 0} \int_I ~dE ~ \int ~d \omega_k ~ \rho(\omega_k)  \Im \langle \varphi, K(\omega_k , E - i \delta ) P_k \varphi \rangle  \nonumber \\
  & \leq & \frac{1}{\pi}  \left( \sup_{\lambda} \rho(\lambda)(1 + \lambda^2) \right)   \lim_{\delta \rightarrow 0}    \int_I ~dE ~ \left| \lim_{\epsilon \rightarrow 0}  \langle \varphi, F_1 ( \epsilon, E - i \delta ) \varphi  \rangle \right| \nonumber   \\
   &\leq & C_W \| \rho \|_\infty  |I| . 
 \eea
By expanding the trace in a basis of ${\rm Ran} ~P_k$, it follows from \eqref{eq:key_identity3} and \eqref{eq:key_identity5} that 
\beq\label{eq:key_identity6}
\E_{\omega_k} \{ {\rm tr} P_k \chi_I(H_{(\omega_k, \omega_k^\perp)}^L ) P_k  \} \leq m C_W \| \rho \|_\infty |I|,
\eeq 
establishing \eqref{eq:spec_ave2}.
\end{proof}


The Wegner estimate for higher-rank Anderson models is now a simple consequence of Proposition \ref{prop:specAve1}.

\begin{theorem}[Wegner Estimate-rank $m$]\label{thm:gWegner}
Let $H_\omega^L$ be the local rank-$m$ Anderson model defined in 
\eqref{eq:localSchr1} with random variables satisfying the above assumptions. 
Then, for any interval interval  $I  \subset \R$, there exists a constant $C_W > 0$, depending on ${\rm supp} ~\rho$, so that
\begin{equation}
\mathbb{P} \{ \tr\chi_I(H_\omega^L)\geq 1 \} \leq \E \{ \tr\chi_I(H_\omega^L) \} \leq C_W \|\rho\|_\infty |\Lambda_L| |I| .
\end{equation}
\end{theorem}

\begin{proof}
As above, the independence of the random variables $\{ \omega_k \}_{k \in \Lambda_{r,L}^*}$ 
allows us to write for any random variable $X = X( \omega)$
$$
\mathbb{E} \{ X \} =\mathbb{E}_{\omega_k^\perp}\mathbb{E}_{\omega_k} \{ X \}, 
$$
for any $k \in \Lambda_{r,L}^*$. 
We note that $\sum_{k \in \Lambda_{r,L}^*} P_k = I_{\Lambda_L}$, the identity on $\ell^2( \Lambda_L)$. This, the Chebychev inequality, the decomposition of the expectation, and spectral averaging imply
\bea\label{eq:wegner2}
\mathbb{P} \{ \tr \chi_I(H_\omega^L)\geq 1 \} & \leq & \mathbb{E} \{ \tr\chi_I(H_\omega^L) \}  \nonumber \\
 & = & \sum_{k \in \Lambda_{r,L}^*} \mathbb{E}_{\omega_{k^\perp}} \mathbb{E}_{\omega_k}  \{ \tr P_k \chi_I(H_\omega^L) P_k \} \nonumber  \\
  & \leq &  C_W \|\rho\|_\infty |\Lambda_L| |I|,
    \eea
  by \eqref{eq:spec_ave2} and the fact that $|\Lambda_{r,L}^*| m = |\Lambda_L|$. 
\end{proof}



\subsection{The generalized Minami estimate}\label{subsec:minami1}

The Minami estimate, originally proven by Minami in \cite{minami1} for lattice random \Schr operators with rank-one Anderson-type potentials, is an upper bound on the probability that $H_\omega^L$ has at least \emph{two} eigenvalues in a small interval. If the eigenvalues of $H_\omega^L$ were independent, then an upper bound would be given by the square of the Wegner estimate, that is 
\beq\label{eq:minamiIndep1}
\mathbb{P} \{ \tr \chi_I(H_\omega^L)\geq  2 \} \leq C_M ( \|\rho\|_\infty |\Lambda_L| |I|)^2 .
\eeq
Minami proved that this upper bound holds even though the eigenvalues are correlated. 
For lattice models, this estimate holds throughout the deterministic spectrum.


\begin{theorem}[Minami Estimate-rank $1$]
\label{thm:Minami}
Let $H_\omega$ be the discrete Anderson model with rank-one perturbations. 
Then, for any interval  $I = [I_-, I_+]$, there exists a constant $C_M > 0$, depending on $I_+$, so
\begin{equation}
\label{eq:Minami0}
\mathbb{P} \{ \tr\chi_I(H_\omega^L)\geq 2 \} \leq C_M ( \| \rho \|_\infty |\Lambda_L| |I| )^2 .
\end{equation}
\end{theorem}

\begin{proof}
Let $H_{\omega_n^\perp,\widetilde{\omega}_n}$ denote the rank-one perturbation of $H_{\omega_n^\perp,\widetilde{\omega}_n}$, that is 
\begin{equation}
H_\omega^L-H_{\omega_n^\perp,\widetilde{\omega}_n}^L = (\omega_n- \widetilde{\omega}_n) | \delta_n \rangle \langle \delta_n | , 
\end{equation}
and let $\widetilde{\omega}_n$ be a random variable, independent of, and with identical distribution to, $\omega_n$.  Weyl's inequality for rank-one perturbations implies that for any interval, $I$,
\begin{equation}
\label{eq:Weyl}
|\tr \chi_I(H_\omega^L) - \tr \chi_I(H_{\omega_n^\perp,\widetilde{\omega}_n}^L) | \leq 1
\end{equation}
We then proceed in a similar manner to the proof of the Wegner estimate, and use the trick above to get a product of independent random variables,
\begin{align}
\mathbb{P} \{ \tr\chi_I(H_\omega^L)\geq 2 \} & \leq \mathbb{E} \{ \tr\chi_I(H_\omega^L)(\tr\chi_I(H_\omega^L)-1) \}  \\
\label{ch1:eq:MinPf2}
& = \sum_{n\in \Lambda_L} \mathbb{E}_{\omega_n^\perp} \mathbb{E}_{\omega_n} \{ \langle \delta_n, \chi_I(H_\omega^L)\delta_n \rangle (\tr\chi_I(H_\omega^L)-1) \}   \\
\label{ch1:eq:MinPf3}
& \leq \sum_{n\in \Lambda_L} \mathbb{E}_{\omega_n^\perp,\widetilde{\omega}_n} \left\{  \mathbb{E}_{\omega_n} \{ \langle \delta_n, \chi_I(H_\omega^L)\delta_n \rangle \right\}   ~(\tr\chi_I(H_{\omega_n^\perp,\widetilde{\omega}_n}^L)    \\
&\leq \sum_{n\in \Lambda_L} \mathbb{E}_{\omega_n^\perp,\widetilde{\omega}_n} \{  \tr\chi_I(H_{\omega_n^\perp,\widetilde{\omega}_n}^L)  \} ( \pi \|\rho\|_\infty |I|) \\
&\leq (\pi \|\rho\|_\infty |\Lambda| |I|)^2
\end{align}
To go from \eqref{ch1:eq:MinPf2} to \eqref{ch1:eq:MinPf3}, we used  \eqref{eq:Weyl}, which importantly, is true for any $\omega_n,\widetilde{\omega_n}\in \R$.
\end{proof}

A key ingredients of this proof of the Minami estimate for rank-one perturbations is the eigenvalue shifting bound \eqref{eq:Weyl}. This allows the replacement of the term $(\tr\chi_I(H_\omega^L)-1)$ in the proof. When the perturbation is rank $m$ as in our model, we replace $1$ by $m > 1$ but this causes a change in the statement of the estimate. We refer to this as a generalized Minami estimate.

We note that the estimate for rank-one perturbations \eqref{eq:Weyl} extends to rank-$m$ perturbations.  A perturbation of rank $m$ can move at most $m$ eigenvalues. So if $\omega_k$ is varied to $\widetilde{\omega}_k$, we have 
\begin{equation}
\label{eq:Weyl2}
|\tr \chi_I(H_{(\omega_k^\perp, \omega_k)}^L) - \tr \chi_I(H_{(\omega_k^\perp,\widetilde{\omega}_k)}^L) | \leq m
\end{equation}

\begin{theorem}[Generalized Minami Estimate-rank $m$]
\label{thm:gMinami}
Let $H_\omega$ be the discrete Anderson model with rank-$m$ perturbations. 
Then, for any interval  $I = [I_-, I_+]$, there exists a constant $C_M >0$ depending on $I_+$ and the support of $\rho$, so that
\begin{equation}
\label{eq:gMinami}
\mathbb{P} \{ \tr\chi_I(H_\omega^L)\geq m+1 \} \leq C_M ( \| \rho \|_\infty |\Lambda_L| |I| )^2 .
\end{equation}
\end{theorem}

\begin{proof}
For an interval $I \subset \R$, we define a set $S_m (I)$ by
$$
S_m(I)  := \{ \omega  ~|~  \tr\chi_I(H_\omega^L)\geq m+1 \}.
$$
We need to estimate 
\beq\label{eq:g_minami1} 
\mathbb{P} \{ \tr\chi_I(H_\omega^L)\geq m+1 \} \leq \frac{1}{m+1} \mathbb{E} \{ \tr\chi_I(H_\omega^L) ( \tr\chi_I(H_\omega^L) -  m ) \chi_{S_m(I)} \} ,
\eeq
where $ \chi_{S_m(I)}$ is the characteristic function of the set $S_m(I)$. This is necessary so the right side of \eqref{eq:g_minami1} is nonnegative. As in the proof of Proposition \ref{prop:specAve1}, we write 
\begin{align}\label{eq:min2}
\lefteqn{ \mathbb{P} \{ \tr\chi_I(H_\omega^L)\geq m+1 \} } \nonumber \\
& \leq \mathbb{E} \{ \tr\chi_I(H_\omega^L)(\tr\chi_I(H_\omega^L)-m) \chi_{S_m(I)} \}  \\
& =   \sum_{k \in \Lambda_{r,L}^*} \mathbb{E}_{\omega_k^\perp} \mathbb{E}_{\omega_k} 
\{   ( {\rm tr} P_k E_{H_\omega^L}(I) P_k ) (\tr   \chi_I(H_\omega^L)-m) \chi_{S_m(I)} \} .
\end{align}
Using the rank-$m$ eigenvalue perturbation inequality \eqref{eq:Weyl2}, with a random variable $\widetilde{\omega}_k$, distributed identically and independently from $\omega_k$, the sum on the last line of \eqref{eq:min2} may be bounded as 
\bea\label{eq:minami2}
\lefteqn{ \sum_{k\in \Lambda_{r,L}^*} \mathbb{E}_{(\omega_k^\perp,\widetilde{\omega}_k)} 
\left\{   \mathbb{E}_{\omega_k} \{ {\rm tr} P_k E_{H_\omega^L}(I) P_k  \} 
\tr \chi_I(H_{(\omega_k^\perp,\widetilde{\omega}_k)}^L ) \right\} } \nonumber 
   \\
&\leq & \sum_{k \in \Lambda_{r,L}^*} \mathbb{E}_{(\omega_k^\perp,\widetilde{\omega}_k)} 
\{  \tr \chi_I ( H_{(\omega_k^\perp,\widetilde{\omega}_k)}^L)  \} ( C_W \| \rho \|_\infty |I| m) .
\eea
Applying the Wegner estimate \eqref{eq:minami1}, and substituting into \eqref{eq:min2}, we obtain the bound
\beq\label{eq:minami3}
\mathbb{P} \{ \tr\chi_I(H_\omega^L)\geq m+1 \}  \leq  C_M ( \| \rho \|_\infty |\Lambda| |I|)^2 ,
\eeq
proving the theorem.  
\end{proof}

%



\section{Removal of eigenvalue degeneracies}\label{sec:evDeg1}
\setcounter{equation}{0}

We begin by studying the deterministic situation when the spectrum of the local Hamiltonian $H_\omega^L$ contains an isolated cluster $I_\epsilon$ of $n$ eigenvalues for a fixed configuration $\omega_0$. 
The first step of the proof is to prove the existence of a configuration $\widehat{\omega}$, close to $\omega_0$,  for which the spacing between \emph{some pair} of successive eigenvalues $E_k$ and $E_{k+1}$ is bounded from below by a positive constant.  Of course, an arbitrary perturbation need not split the degeneracy of eigenvalues, but under suitable conditions, we prove that there is another configuration $\widehat{\omega}$, near $\omega_0$, so that at least one consecutive pair of eigenvalues satisfies $E_{k+1}^{\widehat{\omega}_0} - E_k^{\widehat{\omega}_0} > 0$. Here, we list the eigenvalues in increasing order, including multiplicity.  Furthermore, the new family of $n$ eigenvalues in $I_\epsilon$ for $H_{\widehat{\omega}}^L$ remains suitably separated from the rest of the spectrum of $H_{\widehat{\omega}}^L$.


The question here is how can we guarantee that there is a configuration $\widehat{\omega}$ so that the degeneracy is removed at least between two consecutive eigenvalues. The key idea of Dietlein and Elgart \cite{de} is to use the kinetic energy.  
The kinetic energy plays an important role in forcing the nondegeneracy of at least one pair of consecutive eigenvalues in $I_\epsilon$. This idea is implemented by employing Dirichlet-Neumann decoupling. Let $H_0^{\Lambda_r(k),N}$ and $H_0^{\Lambda_r(k),D}$ be the restriction of $H_0$ to $\Lambda_r(k)$, as in the decomposition \eqref{eq:decomp1},  with Neumann or Dirichlet boundary conditions, as described in Appendix \ref{app3:bc_free1}. For these operators, the Dirichlet-Neumann decoupling is the inequality :
\beq\label{eq:KElb1}
\bigoplus_{k\in \Lambda^*_{r,L}} H_0^{\Lambda_r(k), N} \leq H_0^L \leq \bigoplus_{k \in \Lambda^*_{r,L}} H_0^{\Lambda_r(k),D} . 
\eeq
We refer the reader 
to \cite[Section 5.2]{kirsch2007invitation} for a proof of Dirichlet-Neumann decoupling on the lattice.



\subsection{Eigenvalue splitting: Removal of one degeneracy in $I$}

We begin with a deterministic, perturbative result that guarantees, in the simplest case, the reduction of the degeneracy of an $n$-fold degenerate, isolated eigenvalue by one creating a new cluster of eigenvalues for which the degeneracy of any eigenvalue is less than $n$. As mentioned above, we treat both the upper and lower spectral band edges simultaneously. The region of the spectrum for which the splitting argument works is constrained to small neighborhoods of the lower and upper band edges. 
Some of the results of \cite[section 2]{de} on one parameter perturbations used in this section are presented in appendix \ref{app1:FunctAnaly1}. 

\begin{lemma}[Initial step of lifting the degeneracy]
\label{lemma:evSplitting}
Let $\epsilon \in (0, \frac{1}{12})$. We suppose there is an open interval $I \subset \R$, with $|I| < \frac{1}{2}$, and a configuration $\omega_0 \in [0,1]^{\Lambda_{r,L}^*}$ for which
the following hold:
\begin{enumerate}

\item The local Hamiltonian $H_{\omega_0}^L$  has $n$ eigenvalues $E_i^{\omega_0} \in I$, listed in increasing order counting multiplicity, and the interval $I$ is well-separated from $\sigma(H_{\omega_0}^L) \backslash I$, in the sense that
\begin{equation}\label{eq:well_sep1}
{\rm dist}(I, \sigma(H_{\omega_0})\backslash I ) \geq 8\epsilon .
\end{equation}

\item The interval $I$ is located in an edge of the spectrum. Defining the constant $\gamma_{L,n,r}$ by 
\begin{equation}\label{eq:gamma0n}
\gamma_{L,n,r}:= 2 \left(1- \cos \left( \frac{\pi}{r} \right) \right) \left(1 - \frac{1}{n}- \frac{9\sqrt{2}}{\sqrt{nL}r^d} \right).
\end{equation}
we suppose that $I$ satisfies
\beq\label{eq:interval_edge1}
I \subset  [0,\gamma_{L,n,r}] \cup [4d+1-\gamma_{L,n,r},4d+1] . 
\eeq
Condition \eqref{eq:interval_edge1} implies that the  average of the eigenvalues $\overline{E}^{\omega_0} := \frac{1}{n} \sum_{i=1}^n E_i^{\omega_0}$ satisfies
$$
\overline{E}^{\omega_0} \in [0,\gamma_{L,n,r}] \cup [4d+1-\gamma_{L,n,r},4d+1]  . 
$$


\end{enumerate}

\noindent
Then, 
there exists a configuration $\widehat{\omega} \in \widehat{Q}_\epsilon := \omega_0 + [- {\epsilon}(1-L^{-(2d+1)}), {\epsilon}(1-L^{-(2d+1)})]^{\Lambda_{r,L}^*}$, 
and an integer $1 \leq k \leq n-1$ so that
\beq\label{eq:separation12}
 | E_{k+1}^{\widehat{\omega}} -  E_{k}^{\widehat{\omega}}| > 8  \epsilon L^{-(2d+1)}.
\eeq

\end{lemma}

\begin{proof}
1. We assume that a cluster of $n$-eigenvalues of $H_{\omega_0}^L$ in an interval $I$ satisfies conditions (1) and (2). We now suppose that for all ${\omega} \in \widehat{Q}_\epsilon := \omega_0 + [- {\epsilon}(1-L^{-(2d+1)}), {\epsilon}(1-L^{-(2d+1)})]^{\Lambda_{r,L}^*}$
the eigenvalues of $H_L^{{\omega}}$ in $I$ lie close together
\begin{equation}\label{eq:assump_close1}
\sup_{\omega \in \widehat{Q}_\epsilon} \max_{1 \leq i \leq n}  |E_i^\omega-\overline{E}^\omega| \leq 8 \epsilon L^{-(2d+1)} .   
\end{equation}
We prove that assumption \eqref{eq:assump_close1} implies that the average of these eigenvalues must satisfy: 
\begin{equation}
\overline{E}^\omega \in (\gamma_{L,n,r},4d+1 - \gamma_{L,n,r}),
\end{equation}
contradicting the second assumption (2) since $\omega_0 \in \widehat{Q}_\epsilon$. 
We recall that $r \in \N$ and that $L\in r\N$, so that $\Lambda_L$ consists of $(L/r)^d$ subcubes, each containing $m := r^d$ lattice points.  For $k\in r\Z^d$, we simplify the notation and write 
$\Delta^N_k := H_0^{\Lambda_r(k),N}$ and $\Delta_k^D := H_0^{{\Lambda_r(k)}, D}$ be the Neumann and Dirichlet Laplacians restricted to the subcube, $\Lambda_r(k)$, respectively. Denote the set of points, $k\in r\Z ^d$ that index our subcubes by, $\Lambda^*_{r,L}:= r\Z^d \cap \Lambda^L$, and note that the number of such points is 
$|\Lambda^*_{r,L}| = (L / r)^d$.

%


\noindent
2. The idea is to get upper and lower bounds on the average of the eigenvalues $\overline{E}^\omega$ in $I$:
$$
\overline{E}^\omega := \frac{1}{n}\tr \chi_{I}(H_\omega^L) H_\omega^L \chi_{I}(H_\omega^L),
$$ 
using Dirichlet-Neumann bracketing \eqref{eq:KElb1}, and the positivity of the potential.
For ease of notation, we define $P_\omega:=\chi_{I}(H_\omega^L)$, the projection onto the span of the eigenspaces of $H_\omega^L$ associated to eigenvalues in $I$. First, we have
\beq\label{eq:bounds1}
 \tr P_\omega H_0^L    \leq n \overline{E}^\omega= \tr P_\omega H_\omega^L P_\omega \leq  \tr P_\omega H_0^L  + \| V_\omega^L \| \tr P_\omega  ,
\eeq
and, second, applying Dirichlet-Neumann bracketing, we obtain
the upper and lower bounds 
\begin{equation}
\label{eq:DNtrace1}
\sum_{k \in \Lambda^*_{r,L}} \tr P_\omega \Delta_k^N  \leq \tr P_\omega H_\omega^L \leq  \sum_{k \in \Lambda^*_{r,L}} \left\{  \tr P_\omega \Delta_k^D + \|V_\omega^L\| \tr P_\omega  \right\}.
\end{equation}

\noindent
3. In order to analyze contribution of the free kinetic energy to the lower and upper bounds in \eqref{eq:DNtrace1}, we let $R_{k,0}^N$ be the projection onto the eigenspace corresponding to the smallest eigenvalue of $\Delta_k^N$,  which equal to zero. We let $R_{k,max}^D$ be the projection onto the eigenspace corresponding to the largest eigenvalue of $\Delta_k^D$, which is equal to $4d$.  
Both of these spectral projections are rank-one projections. We let  $(R_{k,0}^N)^\perp$ and $(R_{k,max}^D)^\perp$  denote their complementary subspaces in $\ell^2(\Lambda_r(k))$.
Furthermore, the value $\gamma_r:=2(1- \cos(\pi/r))$ is equal to the second smallest eigenvalue of  
$\Delta_k^N$, and $4d-\gamma_r$ is equal to the second largest eigenvalue of $\Delta_k^D$. Both eigenvalues are $d$-fold degenerate (see Appendix \ref{app3:bc_free1} for a summary of facts concerning $\Delta_k^D$ and $\Delta_k^N$).  Therefore, for the Neumann Laplacian $\Delta_k^N$, we have,
\begin{equation}
\label{eq:laplacianBoundsN1}
\Delta_k^N \geq 0\cdot R_{k,0}^N + \gamma_r (R_{k,0}^N)^\perp  
= \gamma_r \chi_{\Lambda_r(k)} - \gamma_r R_{k,0}^N, 
\eeq
since $\chi_{\Lambda_r(k)} = R_{k,0}^N  + (R_{k,0}^N)^\perp$. 
For the Dirichlet Laplacian $\Delta_k^D$, we have
\begin{equation}
\label{eq:laplacianBoundsD1}
\Delta_k^D \leq 4d R_{k,max}^D + (4d-\gamma_r) (R_{k,max}^D)^\perp 
 = (4d-\gamma_r)\chi_{\Lambda_r(k)} + \gamma_r R_{k,max}^D.
\end{equation}
Combining $\eqref{eq:DNtrace1}$ with the lower bound $\eqref{eq:laplacianBoundsN1}$, we obtain
\begin{equation}\label{eq:meanLB1}
\begin{split}
n\overline{E^\omega} = \tr P_\omega H_\omega^L &  \geq
 \gamma_r \sum_{k\in\Lambda_{r,L}^*} \left\{ \tr P_\omega \chi_{\Lambda_r(k)}P_\omega -  \tr P_\omega R_{k,0}^N P_\omega  \right\} \\
&=  \gamma_r \sum_{k\in\Lambda_{r,L}^*} \left\{ \tr P_\omega \chi_{\Lambda_r(k)}P_\omega -  \tr P_\omega  \chi_{\Lambda_r(k)} P_\omega R_{k,0}^N \right\}  \\
&\geq \gamma_r \sum_{k\in\Lambda_{r,L}^*}  \left\{ \tr P_\omega \chi_{\Lambda_r(k)}P_\omega -  \|P_\omega \chi_{\Lambda_r(k)}P_\omega\|  \tr R_{k,0}^D \right\} \\
&=\gamma_r \tr P_\omega - \sum_{k\in\Lambda^*_{r,L}} \gamma_r \|P_\omega \chi_{\Lambda_r(k)}P_\omega\| ,
\end{split}
\end{equation}
since $\tr R_{k,0}^D =1$. 
Similarly, combining $\eqref{eq:DNtrace1}$ with the upper bound $\eqref{eq:laplacianBoundsD1}$,
we get,
\begin{equation}\label{eq:meanUB1}
\begin{split}
n\overline{E^\omega} = \tr P_\omega H_\omega^L &\leq n\|V_\omega\| + \sum_{k\in\Lambda^*_{r,L}} (4d-\gamma_r)\tr P_\omega \chi_{\Lambda_r(k)}P_\omega + \gamma_r \tr P_\omega R_{k,max}^D \\
&\leq n + \sum_{k\in\Lambda^*_{r,L}} \left\{ (4d-\gamma_r)\tr P_\omega \chi_{\Lambda_r(k)}P_\omega + \gamma_r \|P_\omega \chi_{\Lambda_r(k)} P_\omega \| \tr R_{k,max}^D \right\} \\
&= n + \sum_{k\in\Lambda^*_{r,L}} \left\{  (4d-\gamma_r)\tr P_\omega \chi_{\Lambda_r(k)}P_\omega + \gamma_r \|P_\omega \chi_{\Lambda_r(k)}P_\omega\|  \right\} \\
&=n +  (4d-\gamma_r)\tr P_\omega + \sum_{k\in\Lambda^*_{r,L}} \gamma_r \|P_\omega \chi_{\Lambda_r(k)}P_\omega\|  .
\end{split}
\end{equation}
Dividing by $n$, it follows from \eqref{eq:meanLB1} and \eqref{eq:meanUB1} that the mean $\overline{E^\omega}$ is bounded as 
\begin{equation}
\label{eq:Ebar_bound1}
\gamma_r - \frac{\gamma_r}{n} \sum_{k\in\Lambda^*_{r,L}} \|P_\omega \chi_{\Lambda_r(k)}P_\omega\| \leq \overline{E}^\omega  \leq 1 + 4d-\gamma_r + \frac{\gamma_r}{n} \sum_{k\in\Lambda^*_{r,L}} \|P_\omega \chi_{\Lambda_r(k)}P_\omega\|
\end{equation}

\noindent
4. Let $\alpha_k^\omega := \frac{1}{n}\tr P_\omega\chi_{\Lambda_r(k)}P_\omega$. By the Feynman-Hellmann Theorem, we have that $\alpha_k^\omega = \partial_{\omega_k} \overline{E}^\omega$.
 An application of Lemma \ref{lemma:feynmanHellmann} with 
 $B = \chi_{\Lambda_r(k)}$, $\delta=8\epsilon L^{-(2d+1)}$, and $\epsilon'=\epsilon(1-L^{-(2d+1)})$ guarantees that,
\begin{equation}
|~\alpha_k^\omega - \|P_\omega \chi_{\Lambda_r(k)}P_\omega\|~| < 9\sqrt{2}L^{-d-\frac{1}{2}}
\end{equation}
Summing over $k$, and using that $\sum_k \alpha_k^\omega=1$, result in the bounds,
\begin{equation}
1 - 9\sqrt{2}L^{-d-\frac{1}{2}} \left(\frac{L}{r} \right)^d < \sum_{k\in\Lambda^*_{r,L}}  \|P_\omega \chi_{\Lambda_r(k)}P_\omega\| < 1+9\sqrt{2}L^{-d-\frac{1}{2}} \left(\frac{L}{r} \right)^d
\end{equation}
\noindent
Substituting into $\eqref{eq:Ebar_bound1}$ gives,
\begin{equation}
\gamma_r \left(1 - \frac{1}{n}-\frac{9\sqrt{2}}{\sqrt{nL}r^d} \right) < \overline{E}^\omega < 1+4d - \gamma_r \left(1 - \frac{1}{n}-\frac{9\sqrt{2}}{\sqrt{nL}r^d} \right)
\end{equation}
or,
\begin{equation}
\overline{E}^\omega \in (\gamma_{L,n,r},1+4d-\gamma_{L,n,r}) ,
\end{equation}
proving the desired contradiction.
\end{proof}

\medskip

\begin{remark}
In general, in order to prove a weak Minami estimate for some interval of energy in the deterministic spectrum $\Sigma$ of some specific RSO using the Dietlein-Elgart method,  one should first try to establish the result of  Lemma \ref{lemma:evSplitting} in this interval. This is a key step for the remainder of the arguments in section \ref{sec:evls1} on the EVLS, and in section \ref{sec:minami1} on the weak Minami estimate. 
\end{remark}

\subsection{Eigenvalue splitting: Induction to all eigenvalues in $I$}

In this section, we prove the full initial spacing estimate for the $n$ eigenvalues of $H_\omega^L$ in $I$ by induction on the previous lemma. Technically, we must insure that after each application of  Lemma \ref{lemma:evSplitting} the subclusters of eigenvalues remain separated so that degeneracies removed at the previous step are not re-introduced. 

\begin{theorem}[Induction on the spacing estimate]
\label{thm:evSplitting2}
Let $ \epsilon \in (0, \frac{1}{12})$. We suppose there is an open interval $I \subset \R$, with $|I| < \frac{1}{2}$, and a configuration $\omega_0 \in [0,1]^{\Lambda_{r,L}^*}$ for which the following hold:

\begin{enumerate}
\item The local Hamiltonian $H_{\omega_0}^L$  has $n$ eigenvalues $E_i^{\omega_0}$  in $I$, listed in increasing order counting multiplicity, and the interval $I$ is well-separated from $\sigma(H_{\omega_0}^L) \backslash I$ in the sense of \eqref{eq:well_sep1}, that is,  
$$
{\rm dist}(I, \sigma(H_{\omega_0})\backslash I ) \geq 8\epsilon .
$$

\item The interval $I$ is located in an edge of the spectrum. Recalling the constant $\gamma_{L,2,n}$  defined in \eqref{eq:gamma0n} with $n=2$,
\begin{equation}\label{eq:gamma02}
\gamma_{L,2,r}:= 2 \left(1- \cos \left(\frac{\pi}{r} \right) \right) \left(\frac{1}{2}-\frac{9\sqrt{2}}{\sqrt{L}r^d} \right) ,
\end{equation}
 so $\gamma_{L,2,r} < \gamma_{L,n,r}$, for $n \geq 2$, we assume 
\beq\label{eq:interval_edge2}
I \subset  [0,\gamma_{L,2,r}] \cup [4d+1-\gamma_{L,2,r},4d+1]  , 
\eeq
implying that the  average of the eigenvalues $\overline{E}^{\omega_0} := \frac{1}{n} \sum_{i=1}^n E_i^{\omega_0}$ satisfies
$$
\overline{E}^{\omega_0} \in [0,\gamma_{L,2,r}-\epsilon] \cup [4d+1-\gamma_{L,2,r}+\epsilon,4d+1].
$$


\end{enumerate}
Then, there exists $\widehat{\omega} \in \widehat{Q}_\epsilon := \omega_0 + [- {\epsilon}, {\epsilon}]^{\Lambda_{r,L}^*}$, 
so that
\beq \label{eq:splittingGap}
 \min_{1\leq k \leq n-1} | E_{k+1}^{\widehat{\omega}} -  E_{k}^{\widehat{\omega}}| > 8  \epsilon  L^{-(n-1)(2d+1)}.
\eeq
\end{theorem}

\begin{proof}
This theorem is a result of iterating Lemma $\ref{lemma:evSplitting}$.  Care must be taken to ensure that the hypotheses remain valid after each application of Lemma \ref{lemma:evSplitting}.

\noindent
1.  We first apply the lemma with $\epsilon_1=\epsilon$ and find 
$\omega_1$, $|\omega_1-\omega_0| \leq \epsilon_1(1-L^{-(2d+1)})$, so that for some $1\leq k_1 \leq n-1$,
\begin{equation}
E_{k_1+1}^{\omega_1} - E_{k_1}^{\omega_1} >8\epsilon_1 L^{-(2d+1)}.
\end{equation}

\noindent
2.  Next, let $\epsilon_2:=\epsilon_1 L^{-(2d+1)}$.  Define the groups of eigenvalues from the previous step, $C_1^{\omega_1}=\{ E_1^{\omega_1},\cdots,E_{k_1}^{\omega_1} \}$ and $C_2^{\omega_1}=\{ E_{k_1+1}^{\omega_{1}},\cdots,E_{n}^{\omega_1} \}$.  If $k_1\geq 2$, we apply the lemma to $C_1^{\omega_1}$ with $\epsilon=\epsilon_2$.  If $k_1=1$, we apply the lemma to $C_2^{\omega_1}$.  
We must check that the gaps between groups of eigenvalues are sufficiently large to apply the lemma.  We have $d(C_1^{\omega_1},C_2^{\omega_1})\geq 8\epsilon_2$ as a result of the first iteration of the lemma.  Furthermore, for $i=1,2$
\begin{equation}
d(C_i^{\omega_1}, \sigma(H^L_{\omega_1})\backslash (C_1^{\omega_1}\cup C_2^{\omega_1} ) ) \geq 8\epsilon_1 - 2 |\omega_1-\omega_0| > 6\epsilon_1 > 8\epsilon_2
\end{equation}
Thus, we have $\omega_2$, with $|\omega_2-\omega_1|\leq \epsilon_2(1- L^{-(2d+1)})$ and three groups of eigenvalues, $C_i^{\omega_2}$, $i=1,2,3$.  By the same calculations as in the previous step,
we obtain for $i=1,2,3$
\begin{equation}
d(C_i^{\omega_2},\sigma(H_{\omega_2}^L)\backslash C_i^{\omega_2} ) \geq 8\epsilon_2 L^{-(2d+1)} =:8\epsilon_3
\end{equation}

\noindent
3.  We continue iterating, letting $\epsilon_{j+1}:=\epsilon_j L^{-(2d+1)}$.  After $n-1$ iterations, we find $\omega_{n-1}$, with $|\omega_{n-1}-\omega_{n-2}|\leq \epsilon_{n-1}(1-L^{-(2d+1)}) $ for which each eigenvalue is isolated,
\begin{equation}
d(E_i^{\omega_{n-1}} , \sigma(H^L_{\omega_{n-1}}) \backslash \{E_i^{\omega_{n-1}}\} ) > 8\epsilon_{n-1}L^{-(2d+1)} = 8\epsilon L^{-(n-1)(2d+1)}
\end{equation}
Thus, $\eqref{eq:splittingGap}$ is satisfied for $\widehat{\omega}=\omega_{n-1}$.  Finally, we must check that $|\omega_{n-1}-\omega_0|\leq \epsilon$,
\bea
|\omega_{n-1}-\omega_0| & \leq & \sum_{i=0}^{n-2} |\omega_{i+1}-\omega_i|  \leq \sum_{i=0}^{n-2} \epsilon_{i+1}(1-L^{-(2d+1)}) \nonumber  \\
& = & \epsilon(1-L^{-(2d+1)}) \sum_{i=0}^{n-2}  L^{-i(2d+1)} \nonumber \\
 &  = &  \epsilon(1-L^{-(2d+1)}) \left( \frac{1-L^{-(n-1)(2d+1)}}{1-L^{-(2d+1)}} \right) <\epsilon , 
\eea
proving the theorem.
\end{proof}

To summarize, we have shown that given a configuration $\omega_0$, there is a nearby configuration $\widehat{\omega}$ for which the isolated cluster of $n$ potentially degenerate eigenvalues of the  Hamiltonian $H_{\omega_0}^L$  transforms into an isolated cluster of $n$ nondegenerate eigenvalues of $H_{\widehat{\omega}}^L$, with good control on the size of the separation of the cluster from the rest of the spectrum. 

\section{Eigenvalue level spacing estimate}\label{sec:evls1}


The main result of section \ref{sec:evDeg1} is the existence of good configurations $\omega \in [0,1]^{\Lambda^*_{r,L}}$ for which all $n$ eigenvalues of $H_\omega^L$ in $I$ are nondegenerate. Theorem \ref{thm:evSplitting2} also provides a lower bound on the EVLS of the $n$ eigenvalues originally in the interval $I$. This result, and the Cartan-type lemma of Dietlein-Elgart \cite[Lemma 3.4]{de} (see Appendix \ref{app2:cartan}) are the key ingredients in the proof of the main result of this section, Theorem \ref{thm:spacing}. This theorem provides an upper bound on the probability that the minimum eigenvalue spacing near the band edges is small. We remark again that localization does not play a role in the proof of Theorem \ref{thm:spacing}, see Remark \ref{rmk:loc1}. 

\begin{theorem}[Eigenvalue Level Spacing Estimate]
\label{thm:spacing} Recalling the constant $\gamma_{L, 2,r}$ defined in \eqref{eq:gamma02}, we define the constant
\begin{equation}\label{eq:evls_scale1}
\gamma_{\infty,r}:= \gamma_{\infty,2,r} = (1- \cos(\pi/r)), 
\end{equation} 
and a set of energies near the band edges 
\beq\label{eq:BEinterval1}
I_{sp} :=[0,\gamma_{\infty,r}]\cup[4d + 1 - \gamma_{\infty,r},4d+1] .
\eeq
For any $0<E<\gamma_{\infty,r}$, we define subintervals of $I_{sp}$ by 
\begin{equation}
I_E :=[0,E]\cup [4d+1-E,4d+1] \subset I_{sp} .
\end{equation}
Then, for any  $p>0$, there exist constants $\mathcal{L}_{sp}=\mathcal{L}_{sp}(E,p) >0$ and $C_{sp}=C_{sp}(E,p) > 0$ such that,
\begin{equation}\label{eq:evls1}
\mathbb{P}\{\mathrm{spac}_{I_{E}}(H_\omega^L) <\delta \} \leq C_{sp} L^{2d} |\log \delta|^{-p} ,
\end{equation}
for $L\geq \mathcal{L}_{sp}$ and $\delta\leq \exp(-(\log L)^5)$.
\end{theorem}

\begin{proof}
1. We recall that the uniform rank of the projections $P_k$ in the potential is $m=r^d$. We first decompose $I_{sp}$ into overlapping intervals $\{K_i ~|~ i \in \mathcal{I} \}$ with length $|K_i|=\kappa$, for $\kappa > 0$ to be chosen below, and $|K_{i+1} \cap K_i|\geq \kappa/2$. This implies that $|\mathcal{I}|\leq \lceil |I_{sp}| ~ \left( \frac{2}{\kappa} \right) \rceil$.
Let $K_{i,8\epsilon}:= K_i+[-8\epsilon,8\epsilon]$, for $\epsilon\in (0,1/12)$.  We define the event 
$\Omega_{i,\epsilon}$ by
\begin{equation}
\Omega_{i,\epsilon}:= \{\omega ~|~ \tr \chi_{K_i}(H_\omega^L)\leq m \text{ and } \tr \chi_{K_{i,8\epsilon}\backslash K_i}(H_\omega^L)=0 \}
\end{equation}
The probability of $\Omega_{i,\epsilon}^c$ can be bounded using a Wegner estimate and a generalized Minami estimate \eqref{eq:gMinami} \cite{hk2015} for polymer models,
\begin{equation}\label{eq:prob_comp1}
\begin{split}
\mathbb{P} \{\Omega_{i,\epsilon}^c \} &\leq \mathbb{P} \{ \tr \chi_{K_{i,8\epsilon}\backslash K_i}(H_\omega^L) \geq 1 \} + \mathbb{P} \{ \tr \chi_{K_i}(H_\omega^L) > m \}  \\
&\leq C_W L^d (16\epsilon) + C_{M,m} L^{2d}\kappa^{2} 
\end{split}
\end{equation}
As a consequence of \eqref{eq:prob_comp1}, for $0<\delta<\kappa/2$,
\begin{equation}\label{eq:prob_comp2}
\mathbb{P} \{ \mathrm{spac}_{I_{E}}(H_\omega^L)<\delta \} \leq 16C_W L^d \epsilon |\mathcal{I}| + C_{W,m} L^{2d}\kappa^{2} |\mathcal{I}| + \sum_{i\in \mathcal{I}} \mathbb{P} \{ \{ \mathrm{spac}_{K_i}(H_\omega^L) <\delta\} \cap \Omega_{i,\epsilon} \} .
\end{equation}

\noindent
2. Next, we partition the configuration space, $[0,1]^{\Lambda_{r,L}^*}$ into cubes $Q_j$, $j\in \mathcal{J}$ with side length $2\epsilon$. First, we partition ${\rm Ran} ~ \rho = [0,1]$ as 
$$
[0, 1] = \bigcup_{m=0}^{m^*-1} [2 m \epsilon, 2(m+1)\epsilon] \cup [ 2 m^* \epsilon, 1] 
= \bigcup_{k=1}^{|\Lambda_{r,L}^*|} I_k (\epsilon) ,
$$
where $m^*$ is the smallest integer such that $2 m^* \epsilon \leq 1$, and for some labeling $\{1,2, \ldots, {|\Lambda_{r,L}^*|} \}$ of the points in $\Lambda_{r,L}^*$.  We remark that if we take 
$\epsilon = \frac{1}{k}$, for $k \in \N$, we have $2 m^* \epsilon = 1$.  Consequently, a cube $Q_j \subset 
[0,1]^{\Lambda_{r,L}^*}$ of side length $2 \epsilon$ has the form 
$$
Q_j =I_1 (\epsilon) \times I_2(\epsilon) \times \cdots \times I_ {|\Lambda_{r,L}^*|} (\epsilon) ,
$$
where each interval $I_k(\epsilon) \subset [0,1]$, for $k = 1, \ldots, |\Lambda_{r,L}^*|-1$,  has length $2 \epsilon$. 
The index set $\mathcal{J}$ of the elements $Q_j$ in the partition satisfies $|\mathcal{J}| \leq (\lceil \frac{1}{2\epsilon} \rceil )^{(L/r)^d}$. In the case that $\epsilon = \frac{1}{k}$, for $k \in \N$, we have that $\sum_{j \in \mathcal{J}} \P \{ Q_j \} = 1$. Otherwise, there is overlap at the edges and we bound the sum as follows. Due to the product nature of the probability measure $\P$ on
$[0,1]^{\Lambda_{r,L}^*}$, we compute
$$
\P \{ Q_j \} \leq 2 \epsilon \rho_+ ,
$$
keeping the contribution from one site in $\Lambda_{r,L}^*$, and where $\rho_+ := \sup_{s \in [0,1]} \rho (s)$, the sup of the single-site probability measure $\rho$. As a result,  
\beq\label{eq:partition_sum1}
\sum_{j\in \mathcal{J}} \mathbb{P}  \{Q_j \} \leq 1+ 4\epsilon(L/r)^d \rho_+ .
\eeq

\noindent
3. With this partition, we fix $i,j \in {\mathcal{J}}$ so that $Q_j\cap \Omega_{i,\epsilon}\neq \emptyset$, which is always possible for some $j$ since $\{ Q_j \}$ is a partition and $\Omega_{i,\epsilon} \neq \emptyset$.
For $\omega_{i,j}\in Q_j \cap \Omega_{i,\epsilon}$, the local Hamiltonian $H_{\omega_{i,j}}^L$ satisfies
\begin{equation}
\begin{split}
n_{i,j}:=\tr \chi_{K_i}(H_{\omega_{i,j}}^L) \leq m, \\
{\rm dist} ~ ( (K_i,\sigma(H_{\omega_{i,j}}^L)\backslash K_i ) \geq 8\epsilon  .
\end{split}
\end{equation}  
These conditions on the eigenvalues of $H_{\omega_{i,j}}^L$ guarantee that we can apply 
Theorem $\ref{thm:evSplitting2}$. As a consequence, there exists a configuration 
$\widehat{\omega}_{i,j} \in Q_j$ such that,
\begin{equation}
\mathrm{spac}_{K_{i,\epsilon}}(H^L_{\widehat{\omega}_{i,j}})  \geq 8\epsilon L^{-(n_{i,j}-1)(2d+1)} \geq 8\epsilon L^{-(m-1)(2d+1)}
\end{equation}

\noindent
4. We return to estimate a summand $\mathbb{P} ( \{ \mathrm{spac}_{K_i}(H_\omega^L) <\delta\} \cap \Omega_{i,\epsilon} )$ on the right side of \eqref{eq:prob_comp2}. Decomposing with respect to the sets in the partition $Q_j$, we have 
\bea\label{eq:usingCartan1}
\mathbb{P} \{ Q_j \cap \{ \mathrm{spac}_{K_i}(H_\omega^L) <\delta\} \cap \Omega_{i,\epsilon} \} & \leq & 
\left( \prod_{k\in \Lambda_{r,L}^*} \inf_{(Q_j)_k} \rho \right) | \{ \mathrm{spac}_{K_i}(H_\omega^L) <\delta \} \cap Q_j |  \nonumber \\
 & \leq & ( \rho_-)^{|\Lambda_{r,L}^*|}   | \{ \mathrm{spac}_{K_i}(H_\omega^L) <\delta \} \cap Q_j | ,
\eea  
where $(Q_j)_k$ denotes the interval given by projection of $Q_j$ on the the $k^{\mathrm{th}}$ coordinate. 
In order to estimate the measure of the set on the right in \eqref{eq:usingCartan1}, we apply the Cartan-type lemma in \cite[Lemma 3.4]{de}, stated here as Lemma \ref{lemma:DE_cartan}, with $\delta_0=8\epsilon L^{-(m-1)(2d+1)}$, and obtain the bound,
 \bea\label{eq:usingCartan2}
  | \{ \mathrm{spac}_{K_i}(H_\omega^L) <\delta \} \cap Q_j | & \leq &  C_1 (1+2\epsilon C_\rho)^{(L/r)^d} \mathbb{P}(Q_j) (L/r)^d \exp \left( \frac{-C_2}{r^{2d}} \frac{|\log\delta|}{|\log \delta_0|} \right)  \nonumber \\
&\leq & C_1 (1+2\epsilon C_\rho )^{\left( L/r \right)^d} \mathbb{P}(Q_j) \left( \frac{L}{r} \right)^d  \nonumber \\ 
& & \times \exp \left( \frac{-C_2}{r^{2d}} \frac{|\log\delta|}{|\log(8\epsilon)|+(m-1)(2d+1)|\log L|} \right)  . \nonumber \\
 & & 
\eea
For $0<\delta\leq \exp(-(\log L)^5)$, we choose,
\begin{equation}
\begin{split}
&\kappa := |\log \delta|^{-\alpha} \\
&\epsilon := \exp(-|\log \delta|^{1/4})
\end{split}
\end{equation}
We get,
\begin{equation}
\begin{split}
\mathbb{P} \{ \mathrm{spac}_{I_{sp}}(H_\omega^L)<\delta \} &\leq 16C_W (2|I_{sp}|) L^d \epsilon/\kappa + C_{W,m} (2|I_{sp}|) L^{2d}\kappa^{2-1}  \\ 
&+\sum_{i\in \mathcal{I}} \sum_{j\in \mathcal{J}_i} \mathbb{P} \{ \{ \mathrm{spac}_{K_i}(H_\omega^L) <\delta\} \cap \Omega_{i,\epsilon} \cap Q_j \} \\
&\leq 16C_W (2|I_{sp}|) L^d \epsilon/\kappa + C_{W,m} (2|I_{sp}|) L^{2d}\kappa  \\ 
&+\kappa^{-1} C_1 (1+2\epsilon C_\rho )^{(L/r)^d} (L/r)^d \exp \left( -C_{3,r} \frac{|\log\delta|}{|\log\epsilon|+|\log L|} \right) \\
&\leq C' L^d \exp(-|\log \delta|^{1/4})|\log \delta|^\alpha + C'' L^{2d} |\log \delta|^{-\alpha} \\
&+ C''' L^d |\log \delta|^{\alpha} \exp \left( -C_{3,r} |\log \delta|^{1/2} \right)  .
\end{split}
\end{equation}
Using that $\delta\leq \exp(-(\log L)^5)$, for any $p>0$, we can choose $\alpha$ and $\mathcal{L}_{sp}$ so that,
\begin{equation}
\mathbb{P} \{ \mathrm{spac}_{I_{sp}}(H_\omega^L) <\delta \} \leq C_{sp} L^{2d} |\log \delta|^{-p}.
\end{equation}
This completes the proof of estimate \eqref{eq:evls1}.
\end{proof}

\begin{remark}\label{rmk:loc1}
\begin{enumerate}
 \item We mention the question of the role of localization in the EVLS theorem. In \cite{de}, Theorem 2.1 is a version of the EVLS estimate proven in the localization regime with an improved probability estimate:
 $$
 L^{2d} e^{- |\log \delta|^{\frac{1}{9d}}}.
 $$
This improved probability estimate does not improve the Minami estimate.  To simplify the presentation, we have not included the proof of this improved EVLS estimate, and refer the reader to the proof of Theorem 2.1 in \cite{de}. 

\item  Another simplification of the higher-rank lattice model occurs in the estimation of the probability of $\Omega_{i,\epsilon}^c$ in \eqref{eq:prob_comp1}. The usual Wegner estimate and the generalized Minami estimate, Theorem \ref{thm:gMinami}, provide the necessary bound. For continuum models, Dietlein-Elgart require a refined Wegner estimate \cite[Lemma 4.4]{de}. For any fixed $E$, and constants $\theta, \eta \in (0,1)$, there are constants $c_{\theta, E}, C_{\eta,E} > 0$, such that for any interval $I \subset (-\infty,E]$, 
\beq\label{eq:de_wegner1}
\P \{{\rm tr} \chi_I (H_\omega^L) > c_{\theta, E} |I|^{-\theta} \} \leq C_{\eta, E} |I|^{2 - \eta} .
\eeq
The proof of this estimate requires the spectral shift function and bounds on the expectation of its $L^p$-norm, for $0 < p < 1$.

\end{enumerate}
 
 \end{remark}


\section{The weak Minami estimate}\label{sec:minami1}
\setcounter{equation}{0}

In this section, we prove a weak Minami estimate as a consequence of the eigenvalue level spacing estimate.
Unlike the Minami estimate for the Anderson model on the lattice $\Z^d$ with rank one projectors that holds at all energies in the deterministic spectrum, the Minami estimate of Dietlein-Elgart holds only in small intervals near the lower and upper band edges.

The Minami estimate for the rank-one Anderson model on $\Z^d$ was first proved by Minami in \cite{minami1}. Other proofs have appeared in \cite{bhs,cgk1,gv}. The rank-one nature of the projections seems to be crucial for these proofs. In \cite{hk2015}, a generalized Minami estimate, Theorem \ref{thm:gMinami}, was proved for the higher-rank model discussed here. This estimate has the form
\beq\label{eq:gen_minami1}
\P \{ {\rm Tr} \chi_I (H_\omega^L ) \geq m \} \leq C_M ( |\Lambda_L| |I|)^2 ,
\eeq
for a constant $C_M > 0$. Using this estimate, Hislop and Krishna proved that the LES is a compound point process with L\'evy measure supported on the discrete set $\{ 1, 2, \ldots, m \}$ for energies in the localization regime. In section \ref{sec:PPP1}, we will show that the weak Minami estimate in Theorem \ref{thm:minami1} is sufficient to refine this and show that the process is actually a Poisson point process with an intensity measure $n(E) ~ds$.

\begin{theorem}[Weak Minami Estimate]
\label{thm:minami1}
Let $H_\omega^L$ and $I_{\mathrm{sp}}$ be the same as in Theorem $\ref{thm:spacing}$.  Let $I\subset I_{\mathrm{sp}}$ with $|I|=\delta$.  Fix any $p>0$.  Then, there exists $C>0$, $\mathcal{L}_{\mathrm{sp}}>0$ such that, 
\begin{equation}\label{eq:Minami}
\mathbb{P} \{ \chi_I(H_\omega^L) \geq 2 \} \leq C L^{4d} \delta |\log \delta|^{-p}
\end{equation}
for $L\geq \mathcal{L}_{\mathrm{sp}}$ and $\delta \leq \exp(-(\log L)^{5})$. 
\end{theorem}

\begin{proof}
1. We choose $\mathcal{L}_{\mathrm{sp}}$ large enough so that Theorem $\ref{thm:spacing}$ can be applied, and so that $\delta\leq \exp(-(\log L)^{5d}) \leq L^{-d}/4 \leq |I_{\mathrm{sp}}|$.  Let $\mathcal{I}_1,\mathcal{I}_2$ be intervals so that $I\subset \mathcal{I}_1 \subset \mathcal{I}_2 \subset I_\mathrm{sp}$ and $|\mathcal{I}_1|=\frac{1}{2L^d}$, $|\mathcal{I}_2|=\frac{1}{L^d}$.
With these choices, we cover $\mathcal{I}_2$ with almost disjoint intervals $\{I_i\}_{i=1}^N$ of length $\delta$ (possibly shorter for the intervals intersecting boundaries of $\mathcal{I}_2$) so that $I=I_{i_0}$ for some $i_0 \leq N$ and $\frac{1}{L^d \delta} \leq N \leq \frac{1}{L^d \delta}+2$ .  We compute,
\begin{equation}\label{eq:minami1}
\mathbb{P} \{\tr \chi_I(H_\omega^L)\geq 2 \} \leq \sum_{j=1}^{L^d} \mathbb{P} \{ \mathrm{spac}_{\mathcal{I}_1}(H_\omega^L)\leq \delta \text{ and } E_{\omega,j} \in I \}.
\end{equation}

\noindent
2. For an interval $J \subset \R$, we define the events,
\begin{equation}
\Omega_{i,j}^J := \{ \mathrm{spac}_J(H_\omega^L) < \delta \text{ and } E_{\omega,j} \in I_i \} ,
\end{equation}
and concentrate on estimating $\mathbb{P} \{ \Omega_{i,j}^{\mathcal{I}_1} \}$.
The following key estimate \eqref{eq:minamiKeyEstimate} can be thought of as the uniformity of $\mathbb{P} \{\Omega_{i,j}^{\mathcal{I}_1} \}$ with respect to $i$.  With $\kappa:={(1+L^{-d})}^{-1}$, we  define the deformed Hamiltonian, 
$$
H_{\omega,\kappa}^L := H_\omega^L - (1-\kappa)H_0^L. 
$$
We will also use the notation $E_{\omega,j}^\kappa$ to denote the eigenvalues of $H_{\omega,\kappa}^L$, listed in ascending order.  We claim that for some $C_\rho>0$, and any $1\leq i \leq N$,
\begin{equation}
\label{eq:minamiKeyEstimate}
\mathbb{P} \{ \Omega_{i_0,j}^{\mathcal{I}_1}  \} \leq C_\rho \mathbb{P} \{ \mathrm{spac}_{\mathcal{I}_2}(H_{\omega,\kappa}^L)<\delta  \text{ and } E_{\omega,j}^\kappa \in \kappa I_i \} ,
\end{equation}
recalling that $i_0$ is fixed so that $I=I_{i_0}$.
Assuming $\eqref{eq:minamiKeyEstimate}$, we sum over it over $1 \leq i \leq N$, to obtain: 
\begin{equation}
N \mathbb{P} \{\Omega_{i_0,j}^{\mathcal{I}_1} \} \leq \mathbb{P} \{ \mathrm{spac}_{\mathcal{I}_2}(H_{\omega,\kappa}^L)<\delta \}. 
\end{equation}
So, applying Theorem $\ref{thm:spacing}$ and summing over $j$,  yields the desired estimate,
\begin{equation}
\begin{split}
\mathbb{P} \{ \chi_I (H_\omega^L)\geq 2 \} &\leq \sum_{j=1}^{L^d} \mathbb{P} \{\Omega_{i_0,j}^{\mathcal{I}_1} \} \\
& \leq \sum_{j=1}^{L^d} \frac{1}{N} (C L^{2d} |\log \delta|^{-p}) \\
& \leq C L^{4d} \delta |\log \delta|^{-p}
\end{split}
\end{equation}

\noindent
3. To finish the proof, we must prove \eqref{eq:minamiKeyEstimate}.  This is an application of the basic observation that a shift in energy is equivalent to a shift in the potential provided there is a covering condition. Here, we are in the simplest case that $\sum_{k \in \Lambda_{r,L}^*} V_k = \chi_{\Lambda_L} = \mathrm{Id}_{\ell^2(\Lambda_L)}$, so that,
\begin{equation}\label{eq:covering1}
H_{\omega+\tau}^L = H_\omega^L + \tau \chi_{\Lambda_L} .
\end{equation}
We proceed by working with integrals over the probability space $[0,1]^{\Lambda_{r,L}^*}$. We first make the change of variables, $\omega_k \mapsto \omega_k + \eta_i$, where $\eta_i := d(I,I_i)+\delta$ is the distance from the center of $I$ to the center of $I_i$,
\begin{equation}
\label{eq:keyEstProof1}
\begin{split}
\mathbb{P} \{ \Omega_{i_0,j}^{\mathcal{I}_1} \} &= \int_{[0,1]^{\Lambda_{r,L}^*}} \chi_{\Omega_{i_0,j}^{\mathcal{I}_1}}(\omega) \prod_{k\in \Lambda_{r,L}^*} \rho(\omega_k)~\mathrm{d}\omega_k \\
&\leq \int_{[\eta_i,1+\eta_i]^{\Lambda_{r,L}^*}} \chi_{\Omega_{i,j}^{\mathcal{I}_2}}(\omega) \prod_{k\in \Lambda_{r,L}^*} \rho(\omega_k-\eta_i)~\mathrm{d}\omega_k 
\end{split}
\end{equation}
Next, we make another change of variables $\omega \mapsto \kappa \omega$ in the last integral in \eqref{eq:keyEstProof1}. The purpose of this change is to return the region of integration to one contained in $[0,1]^{\Lambda_{r,L}^*}$. Consequently, we have,
\begin{equation}
\begin{split}
\label{eq:keyEstProof2}
\mathbb{P} \{ \Omega_{i_0,j}^{\mathcal{I}_1} \}  &\leq \kappa^{-(\frac{L}{r})^d} \int_{[\kappa \eta_i,\kappa(1+\eta_i)]^{\Lambda_{r,L}^*}} \chi_{\Omega_{i,j}^{\mathcal{I}_2}}(\kappa^{-1}\omega) \prod_{k\in \Lambda_{r,L}^*} \rho(\kappa^{-1}\omega_k-\eta_i)~\mathrm{d}\omega_k  \\
&\leq \kappa^{-(\frac{L}{r})^d} \int_{[0,1]^{\Lambda_{r,L}^*}} \chi_{\Omega_{i,j}^{\mathcal{I}_2}}(\kappa^{-1}\omega) \prod_{k\in \Lambda_{r,L}^*} \rho(\kappa^{-1}\omega_k-\eta_i)~\mathrm{d}\omega_k 
\end{split}
\end{equation}
Now, we note that,
\begin{equation}
\frac{1}{\kappa}\omega \in \Omega_{i,j}^{\mathcal{I}_2} \iff \mathrm{spac}_{\kappa\mathcal{I}_2}(H_\omega^{L,\kappa})<\kappa \delta \text{ and } E_{\omega,j}^\kappa \in \kappa I_i
\end{equation}
The Lipschitz continuity of $\rho$ gives 
\begin{equation}\label{eq:keyEst11}
\rho(\kappa^{-1}\omega_k-\eta_i) \leq \rho(\omega_k)+2\mathcal{K}L^{-d} \leq \rho(\omega_k)(1+2\mathcal{K}L^{-d}\rho_-^{-1}).
\end{equation}
We note that $(\kappa(1+2\mathcal{K}L^{-d}\rho_-^{-1}))^{-(\frac{L}{r})^d} \leq C_\rho$, for some $C_\rho$ independent of $L$. Using this and the inequality \eqref{eq:keyEst11} in \eqref{eq:keyEstProof2} yields the desired bound,
\begin{equation}
\mathbb{P} \{ \Omega_{i_0,j}^{\mathcal{I}_1}  \}  
\leq  C_\rho \mathbb{P} \{ \mathrm{spac}_{\mathcal{I}_2}(H_{\omega,\kappa}^L)<\delta    \text{ and } E_{\omega,j}^\kappa \in \kappa I_i \}, 
\end{equation}
for \eqref{eq:keyEstProof2}. 
\end{proof}

\begin{remark}\label{rmk:covering2}
\begin{enumerate}
\item Step 3 of the proof uses the simple, but crucial, fact that for lattice models $\sum_{k \in \Lambda_{r,L}^*} \delta_k = \chi_{\Lambda_L} = \mathrm{Id}_{\ell^2(\Lambda_L)}$. This condition insures \eqref{eq:covering1}, a key identity in the proof of  \eqref{eq:minamiKeyEstimate}.  For models on $\R^d$, this requires two modifications: 1) a covering condition, $\sum_{k \in \Z^d} u(x-k) \geq \sigma > 0$, 2) a preliminary step for which $\sum_{k \in \Z^d} u(x-k) = 1$ to insure \eqref{eq:covering1}. This condition must then be relaxed which requires estimates on the modified operator $V^{-1/2} H_0 V^{-1/2}$. 
%
%
\item  We note that an alternate proof of the estimate \eqref{eq:minamiKeyEstimate} would allow a relaxation of the covering condition for the continuum model using the quantitative unique continuation principle as established in \cite{RMV} and \cite{klein1}


\medskip

\item
As mention in section \ref{sec:introduction},  another interesting question is how to extend the Minami estimate to more energies in the spectrum and to improve the probability estimate. 
Considering the question of the region of valid energies, perhaps the initial spacing estimate can be
extended to more energies. However, we can see that the result proven here is not optimal. If it were, we would expect that when $r = 1$, we would recover the result for the rank-one Anderson
model. We know that a Minami estimate holds at all energies for that model, and
while parts of the initial spacing estimate break down for the $r = 1$ case, the weak Minami estimate proven here formally only applies on intervals of length 2 at the edge
of the spectrum, independent of dimension.
\end{enumerate}

\end{remark}


\section{Simplicity of the eigenvalues in the localization regime}\label{sec:simplicity1}
\setcounter{equation}{0}

It was conjectured that the eigenvalues of random \Schr operators in the localization regime are simple, that is, nondegenerate. 
The basic heuristic for this is rooted in the idea that eigenvalue degeneracies are due to symmetries of the \Schr operator. Since these symmetries are destroyed by random perturbations, the eigenvalues should be simple (see \cite{nns1} for an exploration of this idea). For lattice models, this was proved by Simon \cite{simon1}, Minami \cite{minami1}, and by Klein and Molchanov \cite{kleinMolchanov}, the later two works using the Minami estimate, in addition to localization. Dietlein and Elgart \cite{de} showed that the level-spacing estimate and part of the Klein-Molchanov argument yield the simplicity of eigenvalues in the region of the localization regime near the bottom of the almost sure spectrum for random \Schr operators on $L^2(\R^d)$. For the higher-rank models described in section \ref{subsec:model1},
the generalized Minami estimate allows one to prove that the multiplicity of eigenvalues in the localization regime is bounded by the uniform rank of the projections. This can be improved using the EVLS estimate of Theorem \ref{thm:spacing}. We sketch the Dietlein-Elgart argument \cite[section 6]{de} to prove the simplicity of eigenvalues for the rank-$m$ model near the band edges where both the EVLS estimate Theorem \ref{thm:spacing} and localization hold. 

Let $\Omega_{loc} \subset \Omega$ be the set of configurations for which there exists $E_{loc} > 0$ such that $[0, E_{loc} ] \cup [4d +1 - E_{loc}, , 4d+1] $ is in the region of pure point spectrum with exponentially decaying eigenfunctions with $\P \{ \Omega_{loc} \} = 1$.

\begin{theorem}\label{thm:simplicity1}
Let $H_\omega$ be the \Schr operator of the uniform rank-$m$ Anderson model described in section \ref{subsec:model1}. For any $0 < E_0 < \min ( \gamma_{\infty,r}, E_{loc} )$, where $E_{loc}$ is  defned above and $\gamma_{\infty,r}$ is defined in \eqref{eq:evls_scale1},  the eigenvalues of 
$H_\omega$ in $I_{E_0} := [0,E_0] \cup [4d+1-E_0 , 4d+1]$ are almost surely simple. 
\end{theorem}

\begin{proof}
1. The first part of the proof is deterministic and follows \cite[Lemma 1]{kleinMolchanov}. It consists of showing that the existence of two or more linearly independent, exponentially decaying eigenvectors of $H_\omega$, with $\omega \in \Omega_{loc}$ fixed, for an eigenvalue $E$, 
implies that the local \Schr operator $H_\omega^L$
has at least two eigenvalues in a suitably scaled interval $I_\epsilon (E) \subset I_{E_0}$ about $E$. 
We set $\epsilon_L := e^{- \sqrt{L}}$.
Let $\varphi_j$, for $j=1,2$, be two orthonormal eigenvectors with $H_\omega \varphi_j = E \varphi_j$. Let $\chi_L$ denote the characteristic function for a cube of side $L$ centered at the origin. Then, we define the local \Schr operator $H_\omega^L := \chi_L H_\omega \chi_L$, and  consider the localized vectors 
$\varphi^L_j := \chi_L \varphi_j$. Due to the exponential decay of the eigenvectors $\varphi_j$, it is easy to check that for all large $L >0$:
\begin{enumerate}\label{eq:approx_ev1}
\item $0 < (1 - \epsilon_L^2)^\frac{1}{2} \leq  \| \varphi_j^L \| \leq 1$ ; 

\medskip

\item $\langle \varphi_1^L , \varphi_2^L \rangle \leq \epsilon_L $ ;

\end{enumerate}
Furthermore, a short calculation shows that 
$$
\| (H _\omega^L - E ) \varphi_j^L \| \leq  \| \chi_L [ - \Delta, \chi_L ] \varphi_j \| \leq \epsilon_L. 
$$
Consequently, the two vectors $\varphi_j^L$ are linearly independent approximate eigenvectors for $H_\omega^L$. As a consequence, for all $L$ large, ${\rm Tr} ( E_{H_\omega^L} (I_\epsilon(E)) \geq 2$.
 
 \noindent
2. As in \cite[section 6]{de}, we now apply Theorem \ref{thm:spacing} on the eigenvalue level spacing with $\delta = e^{- \sqrt{L}}$. This gives a bound on the the probability of small gaps in the spectrum.
Let us define a set $\Omega_\infty$ of configurations by
\beq\label{eq:gap1}
\Omega_\infty := \{ \omega ~|~ {\rm spac}_{I_{E_0}} H_\omega^L   \leq   e^{- \sqrt{L}}, ~{\rm for ~infinitely ~many} ~ 
L \in \N \} . 
\eeq
By Theorem \ref{thm:spacing} with $\delta = e^{- \sqrt{L}}$ and with $p = 4(d+1)$, we have
\beq\label{eq:gap2}
\P \left\{ {\rm spac}_{I_{E_0}} H_\omega^L   \leq   e^{- \sqrt{L}} \right\} \leq \frac{C}{L^2} ,
\eeq
so for $L = N$, this probability is summable. By the Borel-Cantelli Theorem, we have $\P \{ \Omega_\infty \} = 0$. 
We define another set of configuration $\Omega_2$ defined by
$$
\Omega_2 := \{ \omega ~|~  H_\omega ~{\rm has~an ~eigenvalue } ~\widetilde{E} \in I_{E_0} ~{\rm with ~multiplicity} \geq 2 \}.
$$
For any $\omega \in \Omega_{loc} \cap \Omega_2$, if $\widetilde{E}$ is an eigenvalue of $H_\omega$ in $I_{E_0}$, then ${\rm Tr} ( E_{H_\omega^L} (I_{\epsilon_L}( \widetilde{E})) \geq 2$, for all large $L \in \N$, as follows from step 1. Hence, such a configuration $\omega$ belongs to the set $\Omega_\infty \cap \Omega_2$, which is a set of measure zero, so all eigenvalues of $H_\omega$ in $I_{E_0}$ are simple almost surely. 
\end{proof}

\begin{remark}
Klein and Molchanov only require rapid decay of the eigenvectors like $\langle x \rangle^{- \beta}$, for $\beta > \frac{5d}{2}$.  We also note that the simplicity of the eigenvalues follows from localization and the EVLS estimate, whereas Klein and Molchanov used the Minami estimate. 
\end{remark}


\section{The LES is a Poisson point process}\label{sec:PPP1}
\setcounter{equation}{0}

One of the main applications of a Minami estimate is in proving that the local eigenvalue statistics, $\xi_{\omega,E}^L$, converges weakly to a Poisson point process.
In this section, we outline the proof given in \cite{minami1} with some modifications from \cite{cgk2} in order to illustrate that the weaker form of the Minami estimate in Theorem \ref{thm:minami1} is sufficient to prove that the LES is a  Poisson point process with an intensity measure determined by the DOS. 

Let us recall the definition of the local eigenvalue statistics (LES) for a local, random \Schr operator $H_\omega^L$. Let $\{ E_j^L(\omega) \}$ denote the eigenvalues of $H_\omega^L$, including multiplicity. For a fixed energy $E$ in the deterministic spectrum, we form the local point process, or local random measure, on $\R$, by
\beq\label{eq:pt_proc1}
 \xi_{\omega,E}^L (ds) = \sum_{j \in \N} ~\delta ( |\Lambda_L|(E - E_j^L(\omega) )- s))~ds,
 \eeq
 where $\delta(s-E)$ is the delta distribution centered on $E \in \R$. The random variables $|\Lambda_L|(E - E_j^L(\omega))$ are the rescaled eigenvalues of $H_\omega^L$ centered at energy $E$. The scaling reflects the fact that the Wegner estimate indicates that the average eigenvalue spacing is of size $|\Lambda_L|$. 
 We are interested in the weak limit of $\xi_{\omega,E}^L$ as $L\rightarrow \infty$. The delta functions capture the eigenvalues in a neighborhood of size $|\Lambda|$ about $E$.

 \begin{theorem}\label{thm:poisson1}
Let $E < \min \{ \gamma_{\infty,r}, E_{loc} \}$, where $\gamma_{\infty, r}$ is defined in \eqref{eq:evls_scale1} and $E_{loc}$ is defined in section \ref{sec:simplicity1}. The LES 
$\xi_{\omega,E}^L$ converges weakly to the Poisson point process on $\R$ with intensity measure $n(E) ~ds$, where $n(E) > 0$ is the density of states for the family $H_\omega$.
\end{theorem}

The basic idea of the proof is that the point process constructed from the limit of appropriately scaled families of \emph{independent} random variables should be Poisson. Although the eigenvalues of $H_\omega^L$ are not independent, they are almost independent in the localization regime in the following sense. Because of localization, most of the eigenvectors are concentrated in much smaller regions of $\Lambda_L$, on a scale $\ell \ll L$. as in the proof of Theorem \ref{thm:simplicity1}, we can construct approximate eigenvectors of $H_\omega^\ell$ so that the corresponding eigenvalues are close to the eigenvalues of $H_\omega^L$. Dividing the cube $\Lambda_L$ into a collection of nonoverlapping subcubes $\Lambda_\ell$, we arrive at a collection of local Hamiltonians $H_\omega^\ell$ whose eigenvalues are independent for different subcubes.


In order to implement this idea, we follow \cite{minami1}. 
We always assume that $\ell, L \in r \Z$, with $\ell = o(L)$. 
We divide $\Lambda_L$ into disjoint boxes of side length $\ell$, so that up to sets of measure zero coming from the boundaries, 
\begin{equation}
\Lambda_L = \bigcup_{j=1}^{M_\ell} \Lambda_{\ell,j} .
\end{equation}
We associate a local Hamiltonian $H_\omega^{\Lambda_{\ell,j}}$ with each region $\Lambda_{\ell,j}$ so that, by construction, the family of random Hamiltonians operators $\{H_\omega^{\Lambda_{\ell,j}}\}$ are independent operators.  We define random point measures for each of these operators in analogy with $\xi_{\omega,E}^L$:
\begin{equation}
\xi_{\omega,E}^{\ell,j}(I) := \tr \chi_{E+|\Lambda_L|^{-1}I}(H_\omega^{\Lambda_{\ell,j}})
\end{equation}
We note that the scaling of the energy interval is that same as for $\xi_{\omega,E}^L$. 
From this array of local point processes $\{ \xi_{\omega,E}^{\ell,j} |~|~ j =1, \ldots , M_\ell \}$, we form the superposition
$$
\zeta_{\omega,E}^L : =\sum_{j =1}^{M_\ell} \xi_{\omega,E}^{\ell,j} .
$$

\begin{proof}
1. The first step of the proof is to show that the two local point processes $\zeta_{\omega,E}^L$ and $\xi_{\omega,E}^L$ have the same limit points as $L\to \infty$.  For appropriately chosen $\ell$, this is a consequence of spectral localization since $E$ is in the localization regime.

\noindent
2. The second step is to show that the family  $\{ \xi_{\omega,E}^{\ell,j} ~|~ j =1, \ldots , M_\ell \}$ forms a uniformly, asymptotically-negligible array (uana).  This requires that processes $\xi_{\omega,E}^{\ell,j}$ be independent for different $j \in \{1, \ldots , M_\ell \}$, which they are by construction. The array must also satisfy the following condition that follows from the Wegner estimate:
\beq\label{eq:uana}
\lim_{L\to \infty} \sup_{j  =1, \ldots , M_\ell}  \mathbb{P} \{ \xi_{\omega,E}^{\ell,j}[I]\geq 1 \} 
=0 ,.\eeq
These two conditions establish that $\{ \xi_{\omega,E}^{\ell,j} ~|~ j =1, \ldots , M_\ell\}$ is a uana. 

\noindent
3. There are well-known necessary conditions on a uana that guarantee the weak convergence of the random process $\zeta_{\omega,E}^L$, constructed from its superposition,  to a Poisson point process. The uana must satisfy a set of three conditions, which, from the theory of point processes, imply weak convergence to a Poisson point process with intensity measure $n(E) ds$, see \cite[Theorem 9.2.V]{DVJ}. The conditions are that for any bounded, fixed interval, $I \subset \R$, the following hold,
\begin{equation}
\label{ch1:DVcond1}
\lim_{L\to \infty} \sup_{j =1, \ldots , M_\ell} \mathbb{P} \{ \xi_{\omega,E}^{\ell,j}(I)\geq 1 \} = 0 ;
\end{equation}
\begin{equation}
\label{ch1:DVcond2}
\lim_{L\to \infty} \sum_{j =1, \ldots , M_\ell} \mathbb{P} \{ \xi_{\omega,E}^{\ell,j}(I)\geq 1 \} = n(E) |I| ;
\end{equation}
\begin{equation}
\label{ch1:DVcond3}
\lim_{L\to \infty} \sum_{j =1, \ldots , M_\ell} \mathbb{P} \{ \xi_{\omega,E}^{\ell,j}(I)\geq 2 \} = 0.
\end{equation}

\noindent
4. The Wegner estimate quickly implies \eqref{ch1:DVcond1}.  Indeed, we obtain
$$
\mathbb{P} \{ \xi_{\omega,E}^{\ell,j}(I)\geq 1 \} \leq \E \{ \xi_{\omega,E}^{\ell,j}(I) \}
\leq C_W \left( \frac{\ell}{L} \right)^d |I|   ,
$$
uniform with respect to $j$. Since $\ell = o(L)$, the result follows. 

\noindent
5. We next show \eqref{ch1:DVcond3} that will follow from the weak Minami estimate of Theorem \ref{thm:minami1}. 
From the definitions of the local processes $\eta_{\omega,E}^{\ell,j}$, we have 
\begin{align}\label{eq:les0}
\sum_j \mathbb{P}\{ \xi_{\omega,E}^{\ell,j}(I)\geq 2 \} & \leq \left(  \frac{L}{\ell} \right)^d  \sup_{j =1, \ldots , M_\ell} \mathbb{P}\{\xi_{\omega,E}^{\ell,j}(I)\geq 2 \}  \nonumber \\
&\leq \left( \frac{L}{\ell} \right)^d \sup_{j =1, \ldots , M_\ell} \mathbb{P} \{ \tr \chi_{E+|\Lambda_L|^{-1}I} (H_\omega^{\Lambda_{\ell,j}}) \geq 2 \}.
\end{align}
We estimate the last probability in \eqref{eq:les0} using the weak Minami estimate. For any $p>0$, there is a constant $C_M >0$, and a length scale  $\mathcal{L}_{\mathrm{sp}}>0$ such that, 
\begin{equation}\label{eq:les1}
 \mathbb{P} \{ \tr \chi_{E+|\Lambda_L|^{-1}I}(H_\omega^{\Lambda_{\ell,j}}) \geq 2 \}
 \leq C_M \ell^{4d} \delta |\log \delta|^{-p}  ,
\end{equation}
for all $L\geq \mathcal{L}_{\mathrm{sp}}$ and $\delta \leq \exp(-(\log L)^{5})$. Choosing $\delta = e^{-L}$ and $p = 2$, we find that 
\begin{equation}\label{eq:les2}
 \mathbb{P} \{ \tr \chi_{E+|\Lambda_L|^{-1}I}(H_\omega^{\Lambda_{\ell,j}}) \geq 2 \}
 \leq C_M \left(  \frac{\ell^{4d}}{L^2} \right)  e^{-L},
\end{equation}
establishing the result. 

\noindent
6. The second condition \eqref{ch1:DVcond2} is a consequence of localization and  standard results on density of states, along with \eqref{ch1:DVcond3}. The sum on the left in \eqref{ch1:DVcond2} may be replaced with an expectation as follows. We have 
 \beq\label{eq:dos3}
 \mathbb{P} \{ \xi_{\omega,E}^{\ell,j}[I]\geq 1 \}= \E \{  \xi_{\omega,E}^{\ell,j}[I] \} -
  \sum_{t \geq 2} \P \{  \xi_{\omega,E}^{\ell,j}[I] \geq t \} ,
  \eeq
as follows from the definition of the expectation and the fact that $\xi_{\omega,E}^{\ell,j}[I]$ is integer-valued. The weak Minami estimate \eqref{eq:Minami} allows us to bound the sum on the right in \eqref{eq:dos3}. First, we have
 \bea\label{eq:dos4}
\sum_{t \geq 2}     \P \{  \xi_{\omega,E}^{\ell,j}[I] \geq t \}  & = &  \sum_{t \geq 2} (t-1)\P \{  \xi_{\omega,E}^{\ell,j}[I] = t \}  \nonumber \\
     & \leq &   \sum_{t \geq 2} t(t-1)\P \{  \xi_{\omega,E}^{\ell,j}[I] = t \}  \nonumber \\
     & \leq & \E \{ \xi_{\omega,E}^{\ell,j}[I] ( \xi_{\omega,E}^{\ell,j}[I] -1) \} \nonumber \\
      & \leq & C \ell^{4d} \delta | \log \delta |^{-p} .
      \eea
 As above, we choose $\delta = e^{-L}$ and $p = 2$, we have
\beq\label{eq:dos5}
\sum_{t \geq 2}     \P \{  \xi_{\omega,E}^{\ell,j}[I] \geq t \}  \leq C \left( \frac{\ell^{4d}}{L^2} \right) e^{-L} .
\eeq
The relation between $\zeta_{\omega,E}^L$ and $\xi_{\omega,E}^L$ discussed in step 1 allow us to to conclude that 
$$
\lim_{L \rightarrow \infty} \E \{ \zeta_{\omega,E}^L [I] \} = \lim_{L \rightarrow \infty} \E \{ \xi_{\omega,E}^L [I] \},
$$
so we may replace  $\E \{ \zeta_{\omega,E}^L [I]$ with $\E \{ \xi_{\omega,E}^L [I] \}$.
Using Stone's formula, we may write
\beq\label{eq:les4}
\E \{ \xi_{\omega,E}^L [I] \} = \frac{1}{\pi} \lim_{\epsilon \rightarrow \infty} \sum_{k \in \Lambda_{L}} \int_{E + \frac{1}{|\Lambda_L|} I}  \E \{ \Im \langle \delta_k , R_{\omega,L} (s + i \epsilon ) \delta_k \rangle \} ~ds.
\eeq 
We next use localization in order to replace estimates involving $H_\omega^L$ with $H_\omega$. 
As in \cite{minami1}, this allows us to prove that for $k \in \Lambda_L$:
\beq\label{eq:loc1}
\E \{ \Im \langle \delta_k , R_{\omega,L} (s + i \epsilon ) \delta_k \rangle  \} = \E \{  \Im \langle \delta_k , R_{\omega} (s + i \epsilon ) \delta_k \rangle  \} + \mathcal{O}(e^{-L}) .
\eeq
We also know that the DOS $n$ for $H_\omega$ is related to the Green's function by the formula
\beq\label{eq:dos6}
\E \{ \Im \langle \delta_k , R_{\omega} (s + i \epsilon ) \delta_k \rangle \}  = \int_\R \frac{\epsilon}{(s-t)^2 + \epsilon^2} ~n(t) ~dt  ,
\eeq
independent of $k \in \Z^d$. Since the DOS is a continuous function, we may evaluete the $\epsilon \rightarrow 0$ limit of the right side of \eqref{eq:dos6} and obtain $\pi n(s)$. Using the ergodicity of the matrix elements in \eqref{eq:dos6}, we have 
\bea\label{eq:dos7}
\lim_{L \rightarrow \infty} \E \{ \zeta_{\omega,E}^L [I] \} &=& \lim_{L \rightarrow \infty} 
{|\Lambda_L|}  
\int_{E + \frac{1}{|\Lambda_L|} I}  ~n(s) ~ds  \nonumber \\
 &=& |I| \left[  \lim_{L \rightarrow \infty}  \left( \frac{|I|}{|\Lambda_L|} \right)^{-1} 
 \int_{E + \frac{1}{|\Lambda_L|} I}  ~n(s) ~ds  \right] \nonumber \\
  & = & |I| n(E),
  \eea
  by the Lebesgue Differentiation Theorem. This proves  \eqref{ch1:DVcond2} and hence the process $\zeta_{\omega,E}^L$ converges weakly to a Poisson point process with intensity measure $n(E) ~ds$. 
\end{proof}


\begin{appendices}

\section{One-parameter perturbations}\label{app1:FunctAnaly1}
\setcounter{equation}{0}

In this section, we recall a key lemma from \cite{de}.  The proof, along with intermediate lemmas used in the proof, can be found in \cite[Section 3]{de}.

This lemma is independent of the specific model we are interested in and relies only on functional analysis. In this appendix, we deal with one-parameter perturbations of a self-adjoint operator $A$. 
Let $A$ be self-adjoint operator on a separable Hilbert space, $\mathcal{H}$ and let $I\subset \R$ be an interval with $|I|\leq \frac{1}{2}$. We assume $A$ has $n$ eigenvalues in $I$ and there exists a gap between $I$ and the rest of the spectrum of $A$. Specifically, for any $\epsilon \in (0,\frac{1}{12})$, we suppose that
\begin{equation}
\begin{split}
&n:=\tr \chi_I(A)<\infty, \\
&\mathrm{dist}(I,\sigma(A)\backslash I) \geq 6\epsilon 
\end{split}
\end{equation}
where ${\rm dist}(A,B)$ denotes the distance between two sets $A$ and $B$ in $\R$. 

Let $B$ be a bounded, self-adjoint operator with $\|B\|\leq 1$ and consider the one-parameter family of operators,
\begin{equation}
A_s:= A+sB
\end{equation}
for $s\in (-\epsilon,\epsilon)$.  
Let $I_\epsilon=I+(-\epsilon,\epsilon)$ and let $\{E_i^s\}_{i=1}^n$ be the eigenvalues of $A_s$ in $I_\epsilon$.  We also denote the average of these eigenvalues, $\overline{E^s}=\frac{1}{n}\sum_{i=1}^n E_i^s$. We let $P_s$ denote the spectral projection for the self-adjoint operator $A_s$ and interval $I_\epsilon$. 

\begin{lemma}\cite[Lemma 3.1]{de} 
\label{lemma:feynmanHellmann}
Let $0 < \delta < \epsilon <\frac{1}{12}$. If the eigenvalues satisfy
\beq\label{eq:spacing1}
\sup_{s \in[-\epsilon, \epsilon]} ~\sup_{i=1, \ldots, n} ~| E_i^s - \overline{E}^s
| \leq \delta,
\eeq
then
\beq\label{fluctuationBdd1}
\| P_s ( B - \partial_s \overline{E}^s) P_s \| \leq 9 \sqrt{ \frac{\delta}{\epsilon} } .
\eeq
\end{lemma}


\section{A Cartan-type lemma and the size of bad configurations}\label{app2:cartan} 
\setcounter{equation}{0}

The perturbation theory of section \ref{sec:evDeg1} proves that under the conditions of Theorem \ref{thm:evSplitting2}, there exists a configuration $\widehat{\omega}$ for which the eigenvalues of 
$H_{\widehat{\omega}}^L$, in a specified interval, are all nondegenerate, and for which the spacings between consecutive pairs of eigenvalues are all  bounded below by  $8 \epsilon L^{(n-1)(2d+1)}$.
On the other hand, we say that a configuration $\omega$ is bad if the spacings for these eigenvalues are uniformly less that some $\delta > 0$. We need an estimate on the probability that these bad configurations occur. 

Dietlein and Elgart \cite{de} utilized a result of Bourgain \cite{bourgain} in order to estimate the probability of bad configurations. The Cartan Lemma established in Bourgain is the following:

\begin{lemma}\label{lemma:cartan1}
Let $F(x_1, \ldots, x_N)$ be a real analytic function on $\Omega := [- \frac{1}{2}, \frac{1}{2} ]^N$ that extends to a analytic function in the disk $D^N$, where $D := \{ z \in \C ~|~ |z| < 1 \}$. 
Furthermore, $F$ satisfies the bound
$$
|F(z_1, \ldots,  z_N )| < 1, ~~~{\rm for} ~~~ (z_1, \ldots, z_N) \in D^N.
$$
Suppose there is a point $a \in \Omega$ so that 
$$
|F(a)| > \epsilon,
$$
for some $0 < \epsilon < \frac{1}{2}$. For any $\delta > 0$, we define the set $E_\delta$ by
$$
E_\delta := \{ x \in \Omega ~|~ |F(x)| < \delta \} .
$$
We then have,
\beq\label{eq:cartanEst1}
| E_\delta | < CN \delta^{ \frac{c}{| \log \epsilon|}},
\eeq
for two constants $C,c > 0$.
\end{lemma}

In an abstract setting, 
they considered a multi-parameter perturbation of an operator $A$, similar to the one-parameter operator studied in appendix \ref{app1:FunctAnaly1}. 
Let $N\in \N$ and $0\leq B_k \leq 1$ be self-adjoint operators for $k\in\{1,\cdots,N\}$ such that $\sum_k B_k \leq 1$.  We consider the $N$-parameter family of operators,
\begin{equation}
(s_1,\cdots,s_N) \mapsto A_s:= A+ \sum_k s_k B_k
\end{equation}
for $(s_1,\cdots, s_N)\in (-\epsilon,\epsilon)^N$.  Let us suppose that $A(s)$ has $n$ eigenvalues $E_j(s)$ in an interval $I \subset \R$.  We define the eigenvalue spacing ${\rm spac}_I(A(s))$ as above by
\beq\label{eq:evls_defn2}
{\rm spac}_I(A_{s}) := \min_{1 \leq i\neq j \leq n} |E_i(s) - E_j(s)| .
\eeq

Bourgain \cite[Lemma 1]{bourgain} used a Cartan-type lemma to prove a Wegner estimate. 
 In \cite[Lemma 3.4]{de}, Dietlein and Elgart apply this to the discriminant constructed from the $n$ eigenvalues in $I$: 
\beq\label{eq:discriminant1}
{\rm disc}_I(A_s) := \prod_{1 \leq i < j \leq n} (E_i(s) - E_j(s) )^2 ,
\eeq
rather than to the spacing function ${\rm spac}_I(A_{s})$, since the spacing function is not analytic in $s$. This is possible since we have 
\beq\label{eq:spacingBound1}
| \{ s \in (-\epsilon, \epsilon)^n ~|~ {\rm spac}_I(A_{s})  < \delta \}| \leq 
| \{ s \in (-\epsilon, \epsilon)^n ~|~   {\rm disc}_I(A_s)  < \delta \} |.
\eeq

The main consequence of Bourgain's Lemma \ref{lemma:cartan1} formulation of the Cartan Lemma is the following result due to Dietlein and Elgart.

\begin{theorem}\cite[Lemma 3.4]{de}\label{lemma:DE_cartan}
Suppose that for some $\delta_0 > 0$, there exists a configuration $s_0 \in (-\epsilon, \epsilon)^N$ so that 
\beq\label{eq:spacingCart1}
{\rm spac}_I(A_{s_0}) > \delta_0 .
\eeq
Then, there exists constantc $C_1, C_2 > 0$, independent of $\epsilon$ and $\delta_0$, so that 
\beq\label{eq:spacingCart2}
| \{ s   \in (-\epsilon, \epsilon)^N ~|~ {\rm spac}_I(A_{s_0}) < \delta \} | \leq C_1 N (2 \epsilon)^N {\rm exp} \left(- \frac{C_2 }{n^2} \left| \frac{\log \delta}{\log \delta_0} \right| \right) ,
\eeq
for all $\delta \in (0, 1)$.  
\end{theorem}

\section{Discrete Laplacians: Boundary conditions, eigenvalues, and eigenvectors}\label{app3:bc_free1}
\setcounter{equation}{0}

We work with the positive Laplacian on $\Z^d$,
\begin{equation}\label{eq:laplacian1}
H_0f(n) = 2df(n) - \sum_{|k-n|=1}f(k)  .
\end{equation}
When we restrict $H_0$ to finite set, $\Lambda\subset \Z^d$, we usually do so in the most natural way by simply truncating the full space operator $H_0^\Lambda := \chi_\Lambda H_0 \chi_\Lambda$, producing what is known as \emph{simple boundary conditions}.
We need different boundary conditions, however, 
in order to prove a lattice version of Dirichlet-Neumann bracketing.  Our definitions of\emph{ Neumann and Dirichlet boundary conditions} are equivalent to the ones given in \cite[Section 5.2]{kirsch2007invitation}, but we give a different formula.  To help define these operators, we define an auxiliary, diagonal operator, $m_\Lambda$, with diagonal terms
\begin{equation}
m_\Lambda(n,n) := \#\{k \in \Z^d ~|~ |n-k|=1, ~k \not \in \Lambda\},
\end{equation}
counting the number of nearest neighbors of $n$ that are not in $\Lambda$, and with off-diagonal terms $m_\Lambda(n,k) = 0, k \neq n$.  

\begin{definition}
Let $H_0$ be the lattice Laplacian defined in \eqref{eq:laplacian1}, and let $\Lambda \subset \Z^d$ be a cube. The restriction of $H_0$ to $\lambda$ with
\begin{enumerate}
\item 
with \textbf{simple boundary conditions}, $H_0^\Lambda$, is defined by
\begin{equation}
H_0^\Lambda := \chi_\Lambda H_0 \chi_\Lambda, 
\end{equation}
\item with \textbf{Dirichlet boundary conditions},  $H_0^{\Lambda,D}$, is defined by
\begin{equation}
H_0^{\Lambda,D} := H_0^\Lambda +m_\Lambda  ,
\end{equation}
\item and with \textbf{Neumann boundary conditions}, $H_0^{\Lambda,N}$, is defined by
\begin{equation}
H_0^{\Lambda,N} := H_0^\Lambda -m_\Lambda.
\end{equation}
\end{enumerate}
\end{definition}

\begin{remarks}
1.) The lattice Laplacian $H_0$ defined in \eqref{eq:laplacian1} is the positive Laplacian so $\sigma(H_0)=[0,4d]$ and the spectrum of each of the above cutoff operators is contained in $[0,4d]$. 
2.) As mentioned above, the Dirichlet and Neumann lattice Laplacians are used in the Dirichlet-Neumann bracketing result \eqref{eq:KElb1}.
\end{remarks}

We enumerate the eigenvalues and eigenfunctions of the Dirichlet and Neumann Laplacians on cubes.
Let $\Lambda_L\subset \Z^d$ be a cube consisting of $L^d$ sites.  Define $\tilde{\Lambda}_L\subset \R^d$ as the union of all cubes of side length $1$ centered at sites in $\Lambda_L$.  For example, if $\Lambda_L=\{1,\cdots,L\}^d$, then $\tilde{\Lambda}_L=[1/2,L+1/2]^d$.  

It can be checked that eigenfunctions of the discrete Laplacian on $\Lambda_L$ with Dirichlet or Neumann boundary conditions are the restriction of eigenvectors of the continuum Laplacian on $\tilde{\Lambda}_L$ with corresponding boundary condition.
We can, therefore, simply enumerate the eigenvectors and eigenvalues of $H_0^{\Lambda_L,D}$ and $H_0^{\Lambda_L,N}$. The eigenvectors have the form:
\begin{equation}
\begin{split}
\psi_{n_1,\cdots,n_d}^{\Lambda_L,D}(k) &= \prod_{i=1}^d\sin \left( \frac{\pi n_i}{L}(k_i - 1/2)\right),  \\
\psi_{m_1,\cdots,m_d}^{\Lambda_L,N}(k) &= \prod_{i=1}^d\cos\left(\frac{\pi m_i}{L}(k_i - 1/2)\right),
\end{split}
\end{equation}
and the corresponding eigenvalues:
\begin{equation}
\begin{split}
\label{eq:bc-evs}
E_{n_1,\cdots,n_d}(H_0^{\Lambda_L,D}) & := 2d - 2\sum_{i=1}^d \cos (\pi n_i/L) , \\
E_{m_1,\cdots,m_d}(H_0^{\Lambda_L,N}) & := 2d - 2\sum_{i=1}^d \cos (\pi m_i/L), 
\end{split}
\end{equation}
where $n_i\in \{1,\cdots,L\}$ and $m_i \in \{0,\cdots,L-1\}$.

\end{appendices}




\end{document}